\documentclass[11pt,reqno,preprint]{article}

\usepackage{jheppub} 

\usepackage{slashed}
\usepackage[shortlabels]{enumitem}
\usepackage{shuffle}
\usepackage{tikz-feynman}

\newcommand{\be}{\begin{equation}} 
\newcommand{\ee}{\end{equation}} 
\newcommand{\tr}{\text{tr}}
\newcommand{\asym}{{\wedge^2F}}
\newcommand{\asyma}{{\wedge^2\bar{F}}}

\newcommand{\eps}{\epsilon}
\newcommand{\lam}{\lambda}
\newcommand{\mycomment}[1]{}
\newcommand{\Ax}{\mathcal{A}^\text{ax}}
\newcommand{\ax}{A^\text{ax}}
\newcommand{\sig}{\sigma}
\newcommand{\floor}[1]{\lfloor#1\rfloor}

\newcommand{\Z}{\mathbb{Z}}

\def\eqn#1{eq.~(\ref{#1})}

\def\spa#1.#2{\left\langle#1\,#2\right\rangle}
\def\spb#1.#2{\left[#1\,#2\right]}
\def\sand#1.#2.#3{%
\left\langle\smash{#1}{\vphantom1}\right|{#2}%
\left|\smash{#3}{\vphantom1}\right]}
\def\sandp#1.#2.#3{%
\left[\smash{#1}{\vphantom1}\right|{#2}%
\left|\smash{#3}{\vphantom1}\right\rangle}
\def\sandpp#1.#2.#3{%
\left[\smash{#1}{\vphantom1}\right|{#2}%
\left|\smash{#3}{\vphantom1}\right\rangle}
\def\sandpm#1.#2.#3{%
\left[\smash{#1}{\vphantom1}\right|{#2}%
\left|\smash{#3}{\vphantom1}\right]}
\def\sandmp#1.#2.#3{%
\left\langle\smash{#1}{\vphantom1}\right|{#2}%
\left|\smash{#3}{\vphantom1}\right\rangle}
\def\sandmm#1.#2.#3{%
\left\langle\smash{#1}{\vphantom1}\right|{#2}%
\left|\smash{#3}{\vphantom1}\right]}
\def\spab#1.#2.#3{\sandmm#1.#2.#3}
\def\spba#1.#2.#3{\sandpp#1.#2.#3}
\def\spaa#1.#2.#3.#4{\sandmp#1.{#2#3}.#4}
\def\spbb#1.#2.#3.#4{\sandpm#1.{#2#3}.#4}
\newcommand{\spdenom}[1]{\spa{1}.{2}\spa{2}.{3}\cdots\spa{#1}.1}

\newbox\charbox
\newbox\slabox
\def\s#1{{      
        \setbox\charbox=\hbox{$#1$}
        \setbox\slabox=\hbox{$/$}
        \dimen\charbox=\ht\slabox
        \advance\dimen\charbox by -\dp\slabox
        \advance\dimen\charbox by -\ht\charbox
        \advance\dimen\charbox by \dp\charbox
        \divide\dimen\charbox by 2
        \raise-\dimen\charbox\hbox to \wd\charbox{\hss/\hss}
        \llap{$#1$}
}}

\def\ksl{\s{k}}
\def\Ksl{\s{K}}

\title{\boldmath Rational QCD loop amplitudes and quantum theories on twistor space}

\author[1]{Lance J. Dixon}
\author[1,2]{and Anthony Morales}

\affiliation[1]{SLAC National Accelerator Laboratory, Stanford University
	\\Stanford, CA 94309, USA}
\affiliation[2]{Physics Department, Stanford University
	\\Stanford, CA 94305, USA}

\emailAdd{lance@slac.stanford.edu}
\emailAdd{ammoral@stanford.edu}

\abstract{We show how curing an anomaly of the twistor uplift of self-dual Yang-Mills theory implies linear relations among one-loop, $n$-gluon, color-ordered subamplitudes in QCD, when all $n$ gluon helicities are positive, or when exactly one is negative. We compute the number of linearly independent subamplitudes as determined by these relations, in terms of unsigned Stirling numbers. Then we use a momentum-twistor parametrization to show that there are no further linear dependencies.}

\begin{document} 
\maketitle
\flushbottom
	
\section{Introduction}
\label{sec:intro}
The study of scattering amplitudes has seen great advances in recent years. On the more applied side, computing higher-point and higher-loop amplitudes in the Standard Model has allowed for more precise comparisons to data collected at particle colliders (see e.g.~refs.~\cite{Heinrich:2020ybq,Andersen:2024czj} and references therein). On the more formal side, amplitudes are fascinating theoretical objects in their own right. They provide insight into the behavior and symmetries of a theory, as well as exhibiting previously unforeseen mathematical structures. Having explicit analytic expressions for amplitudes is paramount for finding such structures, and for better understanding aspects of quantum field theory.

One such structure of interest involves relations among subamplitudes. In gauge theories, subamplitudes are the gauge-invariant kinematic coefficients in a color decomposition of amplitudes into a linear combination of ``color factors'' for a general gauge group.  The color factor is nothing more than a Lie algebra invariant that depends solely on the color charge of each external state, whereas the subamplitude depends only on the momentum and helicity of each state. 

Subamplitudes are ``color-ordered'', that is they depend on the order of their arguments, the numerical labels of the external particles.  ``Relations among subamplitudes'' then refers to how a given ordering of arguments is related to other orderings of the same arguments. At both tree level and one-loop level, subamplitudes for $n$-gluon processes satisfy cyclic and reflection symmetries of their arguments, which follow trivially from Bose symmetry, because color factors obey the same symmetries in a particular basis. These relations imply that the number of independent subamplitudes of an $n$-gluon process is no more than $(n-1)!/2$.  Tree-level subamplitudes satisfy further, more non-trivial relations. From group-theory Jacobi relations one can derive~\cite{DelDuca:1999rs} the Kleiss-Kuijf relations~\cite{Kleiss:1988ne}.  A subset of these relations are called photon-decoupling relations. The Kleiss-Kuijf relations only have constant (integer) coefficients multiplying the subamplitudes.  There are also linear relations among tree subamplitudes that contain momentum-dependent coefficients, the celebrated BCJ relations~\cite{Bern:2008qj, Bjerrum-Bohr:2009ulz, Stieberger:2009hq, Feng:2010my, Cachazo:2012uq}. All together, there are $(n-3)!$ linearly independent $n$-gluon tree subamplitudes.
	
At the loop level, $n$-gluon amplitudes depend on the matter content of the gauge theory.  It is not known whether there are general non-trivial linear relations that hold for generic helicity configurations, like in the tree-level case. Some relations have been discovered for specific theories, specific helicity configurations, or parts of amplitudes. One example occurs in $\mathcal{N}=4$ supersymmetric Yang-Mills theory, where the transcendental parts of the one- and two-loop maximally helicity-violating (MHV) amplitudes obey linear relations that go beyond group theory~\cite{Abreu:2019rpt}.

A particularly nice set of loop amplitudes to search for relations are the one-loop QCD amplitudes where all $n$ gluons have the same helicity (all-plus) and the ones where only one gluon has an opposite helicity (one-minus).   Because the tree amplitudes vanish for these helicity configurations, the one-loop amplitudes must be infrared and ultraviolet finite. Moreover, all unitarity cuts of these one-loop amplitudes vanish in four dimensions, due to the vanishing of the tree-level ones. Therefore, these amplitudes contain no branch cuts or transcendental functions; they are purely rational functions of the kinematics.  They are known analytically for all $n$~\cite{Bern:1993qk,Mahlon:1993si,Bern:2005ji}. The all-plus subamplitudes were initially seen to exhibit a vanishing relation when three of the gluons were converted to photons by summing over a particular set of permutations of the color-ordering~\cite{Bern:1993qk}.  More intricate relations were discovered over a decade ago~\cite{Bjerrum-Bohr:2011jrh}.  The one-minus subamplitudes also obey non-trivial relations~\cite{Bjerrum-Bohr:2011jrh}, including three-photon vanishing, although the explicit forms of the most general such relations have not yet been given.  There is also a general argument for the vanishing of the \emph{rational} part of three-photon $(n-3)$-gluon amplitudes for arbitrary helicities~\cite{Boels:2011tp}.

In this paper, we examine relations among the finite loop subamplitudes from the perspective of twistor theory. Self-dual Yang-Mills (sdYM) theory on twistor space is anomalous, and removing this anomaly renders the $S$-matrix trivial~\cite{Costello:2021bah,Costello:2022upu,Costello:2022wso,Costello:2023vyy}. The all-plus one-loop amplitudes vanish in these anomaly-free theories, and we will derive relations among subamplitudes from this fact. When the anomaly is cured by including fermionic matter in a special representation of $SU(N)$~\cite{Costello:2023vyy}, we will recover the relations conjectured to hold for all $n$ in ref.~\cite{Bjerrum-Bohr:2011jrh}.  We will also see that the three-photon vanishing relations are a subset of more general double-trace relations.  Additionally, we will argue that the one-minus-helicity subamplitudes exhibit the same double-trace relations.

The anomaly can also be cured by introducing a scalar with a fourth-order kinetic term -- nicknamed the ``axion'' -- that couples to an operator that is bilinear in the gauge-field-strength tensor. The axion anomaly cancellation method only works for gauge groups $SU(2)$, $SU(3)$, and the exceptional Lie groups, because it requires that the quartic Casimir be dependent on the quadratic Casimir group invariant. Obtaining relations on subamplitudes from this description requires the use of a color decomposition into symmetrized traces and products of strings of structure constants~\cite{Bandiera:2020aqn}. We will see that the relations obtained from this anomaly cancellation method are equivalent to the ones obtained from the fermionic-matter cancellation.

We will solve the relations implied by these cancellations, in order to determine the number of unconstrained subamplitudes. These numbers will be compared to the linear span of the explicit subamplitudes, as determined using a momentum-twistor parametrization~\cite{Hodges:2009hk,Badger:2013gxa}. We will find that the two numbers agree, i.e.~there are no further linear relations with constant coefficients, for both all-plus and one-minus.  Also, in both cases the number of independent subamplitudes forms a known integer sequence involving the unsigned Stirling
numbers~\cite{Bjerrum-Bohr:2011jrh}.
	
This paper is organized as follows. We define our notation and conventions in section \ref{sec:notation}. In section \ref{sec:colordecomps}, we review tree-level and one-loop color decompositions. Relations following from the anomaly cancellation by including special fermionic matter are examined in section~\ref{sec:matterrelns}, and the ones following from the inclusion of the axion are discussed in section~\ref{sec:axion}. We conclude in section \ref{sec:conclusions}.  We include an appendix on momentum-twistor parametrizations and another on standard factorizations of permutations of a given length. 

\section{Conventions and notation}
\label{sec:notation}
In this section, we define our conventions and notations. The $SU(N)$ Lie algebra generators in an arbitrary representation $R$ are denoted by $t_R^a$.  The generators in the fundamental and adjoint representations receive a special notation,
\begin{equation}
	T^a \equiv t_F^a, \hspace{0.35cm} F^a \equiv t_G^a.
\end{equation}
The adjoint representation matrices have components defined by
\begin{equation}
	(F^b)^{ac} \equiv if^{abc} \equiv F^{abc},
\end{equation}
where the structure constants are real and normalized as
\begin{equation}
	[t_R^a,t_R^b]=if^{abc}t_R^c.
\end{equation}
The Dynkin index $T_F$ of the fundamental representation is taken to be unity,
\begin{equation}
	\tr(T^aT^b)=T_F\delta^{ab}\equiv\delta^{ab}.
\end{equation}
We write the trace of $n$ generators in a representation $R$ compactly as
\begin{equation}
	\tr_R(t^{a_1}t^{a_2}\cdots t^{a_n}) = \tr_R(12\dotsc n).
\end{equation}
If the $R$ is omitted, then it is understood that the trace is taken over the fundamental representation $F$,
\begin{equation}
	\tr(12\dotsc n)\equiv \tr_F(12\dotsc n).
\end{equation}
If $I=(i_1,\dotsc,i_l)$ is some list of length $l$ of positive integers, then we set
\begin{equation}
	\tr_R(I)=\tr_R(t^{a_{i_1}}t^{a_{i_2}}\cdots t^{a_{i_l}}).
\end{equation}
	
We denote the set of all permutations (bijections) of a finite set $X$ by $S(X)$. The symmetric group on $n$ letters is specifically $S_n \equiv S(\{1,\dotsc,n\})$, i.e.~we will differentiate between the symmetric groups of two distinct sets even if they have the same size. For example, we take $S_{n-1}$ and $S(\{2,\dotsc,n\})$ to be distinct. A permutation $\sig\in S_n$ is thought of as a word consisting of letters $\sig(i)$ for $1\leq i\leq n$,
\begin{equation}
	\sig=\sig(1)\sig(2)\cdots\sig(n).
\end{equation}
The reverse order of the word $\sig$ is $\sig^T=\sig(n)\sig(n-1)\cdots\sig(1)$. This means that we can further shorten our notation for traces to
\begin{equation}
	\tr_R(\sig) = \tr_R(\sig(1)\sig(2)\cdots\sig(n))
	=\tr_R(t^{a_{\sig(1)}}t^{a_{\sig(2)}}\cdots t^{a_{\sig(n)}}),
\end{equation}
or, if $\sig\in S_{n-1}$,
\begin{equation}
	\tr_R(\sig n) = \tr_R(\sig(1)\sig(2)\cdots\sig(n-1)n)
	=\tr_R(t^{a_{\sig(1)}}t^{a_{\sig(2)}}\cdots t^{a_{\sig(n-1)}}t^{a_n}).
\end{equation}
Finally, a string of contracted structure constants or, equivalently, a product of adjoint matrices, is denoted by
\begin{equation}
	F^w\equiv F^{w_1w_2\cdots w_k}
	=(F^{w_2}F^{w_3}\cdots F^{w_{k-1}})^{w_1w_k},
\label{eq:Fstringdef}
\end{equation}
where $w$ is some subword of a permutation $\sig\in S_n$ with $k$ letters, for $k\geq3$. In the case that $w$ has length $k=2$, this notation is used to mean
\begin{equation}
	F^w=F^{w_1w_2}=\delta^{w_1w_2}.
\end{equation}

\section{Color bases for one-loop amplitudes}
\label{sec:colordecomps}
	
The vanishing of the one-loop all-plus amplitudes relies on the existence of relations between one-loop color structures among different representations of $SU(N)$. Therefore we need to study the color decomposition of both tree and one-loop amplitudes. We first review the well-known trace basis and DDM basis~\cite{DelDuca:1999rs}. We then describe a newer color basis based on the decomposition of traces into symmetrized traces and structure constants~\cite{Bandiera:2020aqn}.
	
\subsection{Trace basis}
The trace basis for an $n$-point process refers to a decomposition of an amplitude whose color factors are given by some product of traces in the fundamental representation of $SU(N)$. The tree-level decomposition is
\begin{equation}
\label{treetracebasis}
	\mathcal{A}_n^{(0)} = g^{n-2}\sum_{\sig\in S_{n-1}}
		\tr(\sig n)A_n^{(0)}(\sig,n),
\end{equation}
where the $A_n^{(0)}$ are the color-ordered subamplitudes.\footnote{%
Our notation for amplitudes is similar to our notation for traces, e.g.~we use $A_n(\sigma,n)$ when $\sigma$ has $n-1$ arguments (for example, if $\sigma\in S_{n-1}$), and $n$ is not permuted; and we use $A_n(\sigma)$ when $\sigma$ contains all $n$ arguments.}
The one-loop $n$-gluon QCD amplitude in the trace basis is~\cite{Bern:1990ux}
\begin{align}
	\label{tracebasis}
	\begin{split}
		\mathcal{A}_n^\text{(1)} = g^n
		\Bigg[
		&N\sum_{\sigma\in S_{n-1}}\tr(\sig n)
		A_n^{[1]}(\sigma,n)
		\\
		&+\sum_{c=2}^{\lfloor n/2 \rfloor+1}\sum_{\sigma\in S_n/S_{n;c}}\tr(\sigma(1\dots (c-1)))\tr(\sigma(c\dots n))
		A_{n;c}(\sigma)
		\\
		&+n_f\sum_{\sigma\in S_{n-1}}\tr(\sig n)
		A_n^{[1/2]}(\sig,n)
		\Bigg],
	\end{split}
\end{align}
where the $A_{n;c}$ are the double-trace subamplitudes and $n_f$ is the number of quark flavors. The superscript $[j]$ denotes the spin of the particle circulating in the loop. The single-trace subamplitudes $A_n^{[j]}$ are color-ordered. The pure Yang-Mills (YM) amplitude is simply given by taking $n_f=0$. The set $S_{n;c}$ denotes the set of all permutations that leave the double-trace structure invariant.

The subleading double-trace subamplitudes $A_{n;c}$ are obtained from the leading ones $A_n^{[1]}$ through the permutation sum~\cite{Bern:1990ux,Bern:1994zx,DelDuca:1999rs}
\begin{equation}
	\label{Anc}
	A_{n;c}(\alpha,\beta) = (-1)^{|\beta|}
	\sum_{\sigma\in\alpha\shuffle\beta^T}A_n^{[1]}(\sigma),
\end{equation}
where $\alpha=[1,2,\ldots,c-1]$ and $\beta=[c,c+1,\ldots,n]$ are cyclically ordered lists, and $\beta^T$ is the reverse ordering. Square brackets are used (as opposed to parentheses) to serve as a reminder that $\alpha$ and $\beta$ are equivalence classes under cyclic permutations of its arguments, i.e.
\begin{align}
	\alpha &= [1,2,\dotsc,c-1] = \{(1,2,\dotsc, c-1),(2,\dotsc,c-1,1),\dotsc,(c-1,1,\dotsc,c-2)\}
	\\
	\beta &= [c,c+1,\dotsc,n] = \{(c,c+1,\dotsc,n),(c+1,\dotsc,n,c),\dotsc,(n,c,\dotsc,n-1)\}.
\end{align}
The symbol $\alpha\shuffle\beta^T$ denotes the cyclic shuffle product, which is the set of all permutations up to cycles of $\{1,2,\ldots,n\}$ that preserves the cyclic ordering of $\alpha$ and $\beta^T$, while allowing all possible relative orderings of the elements of $\alpha$ with respect to the elements of $\beta^T$. For example, letting $\alpha=[1,2,3]$ and $\beta=[4,5]$, there are 12 possible orders in
\begin{align}
\begin{split}
	\alpha\shuffle\beta^T = 
	\{
	&[1,2,3,5,4],[1,2,5,3,4],[1,5,2,3,4],[1,2,5,4,3]
	\\
	&[1,5,2,4,3], [1,5,4,2,3], [1,2,3,4,5], [1,2,4,3,5]
	\\
	&[1,4,2,3,5], [1,2,4,5,3], [1,4,2,5,3], [1,4,5,2,3]
	\}.
\end{split}
\end{align}
Again, it is understood that the lists within this set are equivalence classes under cyclic permutations of their arguments.

\subsection{DDM basis}
Another one-loop color decomposition exists for adjoint particles circulating in the loop; it is given in terms of strings of structure constants or, equivalently, adjoint matrices. This choice of color basis is often referred to as the DDM basis~\cite{DelDuca:1999rs}. The tree-level amplitude in the DDM basis is
\begin{equation}
\label{treeDDM}
	\mathcal{A}_n^{(0)}
	= g^{n-2} \sum_{\sig\in S(\{2,\dotsc,n-1\})}F^{1\sig n}
	A_n^{(0)}(1,\sig,n),
\end{equation}
where the subamplitudes $A_n^{(0)}$ are the same color-ordered ones as in eq.~\eqref{treetracebasis}. The one-loop QCD amplitude with $n$ external gluons in this basis is
\begin{equation}
	\label{loopDDM}
	\mathcal{A}_n^{(1)}
	= g^n \sum_{\sig\in S_n/R\Z_n} \Bigl[ \tr_G(\sig)A_n^{[1]}(\sig) 
	+ 2n_f\tr(\sig)A_n^{[1/2]}(\sig) \Bigr]\,,
\end{equation}
where the subamplitudes $A_n^{[1]}$ and $A_n^{[1/2]}$ are the same ones that appear in eq.~\eqref{tracebasis}. Note that $S_n/R\Z_n$ means the permutations on $n$ letters modulo those related by cycles ($\Z_n$) and reflections ($R$).

In this basis, the commonly stated reflection identity for both tree and one-loop level amplitudes
\begin{equation}
	\label{refid}
	A_n(1,\dotsc,n)=(-1)^nA_n(n,\dotsc,1)
\end{equation}
is now apparent. It follows from the reflection identity on strings of structure constants, or traces in the fundamental representation,
\begin{eqnarray}
	  \tr_G(12\dotsc n) &=& \sum_a F^{a12\dotsc na} = (-1)^n \sum_a F^{an\dotsc21a} = (-1)^n\tr_G(n\dotsc21) \label{adjtracerefid} \\
         \tr_{\bar{F}}(12\dotsc n) &=& (-1)^n\tr_F(n\dotsc21), \label{fundtracerefid}
\end{eqnarray}
and Bose symmetry. So, there are at most $(n-1)!/2$ linearly independent one-loop subamplitudes, regardless of the helicity structure.

\begin{figure}[h]
	\centering
	\begin{equation}
		\nonumber
		\begin{tikzpicture}[baseline={(0,0)},scale=0.75]
			\begin{feynman}
				\vertex (g1) at (0,-2);
				\vertex (g2) at (0,2);
				\vertex (g3) at (4,2);
				\vertex (g4) at (4,-2);
				\vertex (v1) at (1,-1);
				\vertex (v2) at (1,1);
				\vertex (v3) at (3,1);
				\vertex (v4) at (3,-1);
				\diagram*[edges=gluon]{
					(g1) -- (v1),
					(v2) -- (g2),
					(g3) -- (v3),
					(v4) -- (g4),
					(v1) -- [quarter right] (v4) -- [quarter right] (v3) 
					-- [quarter right] (v2) -- [quarter right] (v1),
				};
			\end{feynman}
		\end{tikzpicture}
		=
		\begin{tikzpicture}[baseline={(0,0)},scale=0.75]
			\begin{feynman}
				\vertex (g1) at (0,-2);
				\vertex (g2) at (0,2);
				\vertex (g3) at (4,2);
				\vertex (g4) at (4,-2);
				\vertex (v1) at (1,-1);
				\vertex (v2) at (1,1);
				\vertex (v3) at (3,1);
				\vertex (v4) at (3,-1);
				\vertex (w1) at (1.25,-0.75);
				\vertex (w2) at (1.25,0.75);
				\vertex (w3) at (2.75,0.75);
				\vertex (w4) at (2.75,-0.75);
				\diagram*{
					(g1) -- [gluon] (v1),
					(v2) -- [gluon] (g2),
					(g3) -- [gluon] (v3),
					(v4) -- [gluon] (g4),
					(v1) -- [quarter left, fermion] (v2) 
					-- [quarter left, fermion] (v3) 
					-- [quarter left, fermion] (v4) 
					-- [quarter left, fermion] (v1),
					(w1) -- [quarter left, anti fermion] (w2) 
					-- [quarter left, anti fermion] (w3) 
					-- [quarter left, anti fermion] (w4) 
					-- [quarter left, anti fermion] (w1),
				};
			\end{feynman}
		\end{tikzpicture}
	\end{equation}
	\caption{Graphical representation of the $SU(N)$ identity $G\oplus1\cong F\otimes\bar{F}$. The diagram on the right is evaluated by summing over all $2^n$ ways to attach $n$ external legs to either ring of the annulus.}
	\label{fig:g=ffbar}
\end{figure}
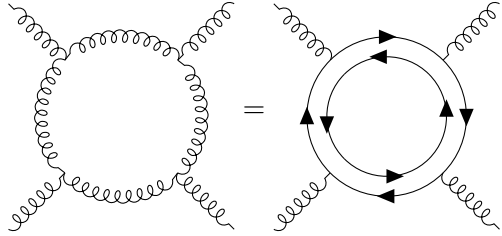

The equivalence of eqs.~\eqref{tracebasis} and \eqref{loopDDM} can be seen through the $SU(N)$ relation $G\oplus1\cong F\otimes\bar{F}$ in terms of traces
\begin{align}
	\label{adjtoF}
	\tr_G(1\ldots n) &= \tr_{F\otimes\bar{F}}(1\ldots n)
	=\sum_{I\subset(1,\ldots,n)}\tr_F(I)\tr_{\bar{F}}(I^c)
	\nonumber
	\\
	&= N\tr_F(1\ldots n) + (-1)^nN\tr_F(n\ldots 1)		
	+\sum_{\emptyset\neq I\subsetneq(1,\cdots,n)} (-1)^{|I^c|}
           \tr_F(I)\tr_F\bigl( (I^c)^T \bigr),~~
\end{align}
where $I^c$ is the complement of the sublist $I$, and we used \eqn{fundtracerefid} in the last line. The notation $I\subset(1,\dotsc,n)$ means that $I$ is a sublist of $(1,\dotsc,n)$ with respect to which $I$ is ordered. This relation has a nice diagrammatic representation in terms of color graphs using the double-line notation, as seen in fig.~\ref{fig:g=ffbar}. As a reminder on the double-line notation, the rule is to sum all $2^n$ ways of attaching the $n$ external lines to either the inner or outer ring of the annulus.

The matter contribution to the color decomposition is more generic than in eqs.~\eqref{tracebasis} and \eqref{loopDDM}. In fact, if the fermions live in some representation $R$ of $SU(N)$, the full amplitude becomes
\begin{equation}
	\label{DDMwithRmatter}
	\mathcal{A}_{n,R}^\text{1-loop}
	= g^n \sum_{\sigma\in S_n/R\mathbb{Z}_n} \Bigl[ \tr_G(\sigma)
	A_n^{[1]}(\sigma) + \tr_R(\sigma)A_n^{[1/2]}(\sigma) \Bigr] \,.
\end{equation}

\begin{figure}[h]	
	\centering
	\begin{tikzpicture}[scale=0.75]
		\begin{feynman}
			\vertex (g1) at (0,-2);
			\vertex (g2) at (0,2);
			\vertex (g3) at (4,2);
			\vertex (g4) at (4,-2);
			\vertex (v1) at (1,-1);
			\vertex (v2) at (1,1);
			\vertex (v3) at (3,1);
			\vertex (v4) at (3,-1);
			\diagram*{
				(g1) -- [gluon] (v1),
				(v2) -- [gluon] (g2),
				(g3) -- [gluon] (v3),
				(v4) -- [gluon] (g4),
				(v1) -- [quarter left, fermion] (v2) 
				-- [quarter left, fermion, edge label=\(R\)] (v3) 
				-- [quarter left, fermion] (v4) 
				-- [quarter left, fermion] (v1),
			};
		\end{feynman}
	\end{tikzpicture}
	\caption{The one-loop color diagram for matter in an arbitrary representation $R$ of $SU(N)$.}
	\label{fig:fermion1loop}
\end{figure}
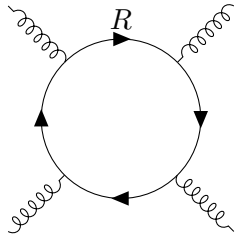

This replacement is permitted for the following reason. Every Feynman diagram can be written as the product of a color factor and a kinematic factor. The Jacobi identity on the color factors can be used to remove color graphs with nontrivial trees attached to the loop~\cite{DelDuca:1999rs}, and thereby rewrite the matter contribution as a sum of permutations of the ``ring'' color diagram in fig.~\ref{fig:fermion1loop}. Because the Jacobi identity is independent of the choice of representation of the fermion loop, we arrive at the same sum over color diagrams, with the same choice of fermion representation with which we began, without affecting the final kinematic factors. That is to say, $A_n^{[j]}$ depends solely on the spin of the particle propagating in the loop, not the representation of the Lie algebra in which it resides.

\subsection{Symmetrized trace basis (Casimir basis)}

While the trace and DDM bases are rather familiar, there is a third type
of color decomposition that is quite useful for discussing the all-plus
amplitude, which is based on symmetrized traces~\cite{Bandiera:2020aqn}.
We will see later that the all-plus amplitude is quite ``diagonal'' in
this basis.  This fact is closely connected to a description
of the independent all-plus amplitudes in terms of a single totally
symmetric quartic vertex and the remaining vertices cubic
and antisymmetric~\cite{Bjerrum-Bohr:2011jrh}.  Finally, this representation
aids the study of implications of the axion cancellation mechanism
in section~\ref{sec:axion}.

Given a simple Lie algebra with generators $t^a_R$ in a representation $R$ satisfying
\begin{equation}
	[t^a_R,t^b_R]=if^{abc}t^c_R,
	\hspace{0.5cm}
	\tr_R( t^at^b ) = T_R~\delta^{ab},
\end{equation}
the color-trace-decomposition problem involves expressing traces of products of generators $\tr_R(t^{a_1}\cdots t^{a_n})$ in terms of products of symmetrized traces, a.k.a.~Casimirs,
\begin{equation}
	d_R^{a_1\dots a_n} = \frac{1}{n!}\sum_{\sig\in S_n}\tr_R(t^{a_{\sig(1)}}\cdots t^{a_{\sig(n)}})
\end{equation}
and structure constants $f^{abc}$ or, equivalently, adjoint generators $F^a$. The first three examples are~\cite{Mafra:2012kh,Bandiera:2020aqn}
\begin{align}
\begin{split}
	\tr_R(12) =&~ d_R^{12},
	\\
	\tr_R(123) =&~ d_R^{123} + \frac{T_R}{2}F^{123},
	\\
	\tr_R(1234) =&~ d_R^{1234} + \frac{T_R}{3}F^{1234} 
			- \frac{T_R}{6}F^{1324} 
			\\
			&+ \frac{1}{2}d_R^{34a}F^{12a}
			+ \frac{1}{2}d_R^{24a}F^{13a} + \frac{1}{2}d_R^{14a}F^{23a},
\end{split}
\end{align}
where we recall the definitions~(\ref{eq:Fstringdef}).
Expressing traces in this manner has the advantage of using Casimir group invariants for the color factors in Feynman diagrams, which allows for efficient generalizations to arbitrary Lie groups/algebras and their representations \cite{vanRitbergen:1998pn}. This representation will help us compare the all-plus one-loop YM amplitudes to the all-plus tree-level graphs with a single axion exchange in section \ref{sec:axion}.

A closed-form solution to this problem was given in ref.~\cite{Bandiera:2020aqn}:
\begin{equation}
	\label{trdecomp}
	\tr_R(12\cdots n) = \sum_{\sig_k\cdots\sig_1=\sig\in S_{n-1}}
		C_{\sig_k}\cdots C_{\sig_1}d_R^{nb_k\dots b_1}
		F^{\sig_kb_k}\cdots F^{\sig_1b_1}.
\end{equation}
Here, $\sig=\sig_k\sig_{k-1}\cdots\sig_1$ denotes the standard factorization of a permutation (or word) $\sig\in S_{n-1}$ into subwords $\sig_1,\dotsc,\sigma_k$ with the properties:
\begin{enumerate}[(1)]
	\item The first letter of $\sig_i$ is greater than the first letter of $\sig_j$ whenever $i>j$.
	\item The first letter of $\sig_i$ is the minimum letter appearing in $\sig_i$.
\end{enumerate}
This factorization is unique, and $\sig$ is viewed as the concatenation of these subwords. Examples include
\begin{equation}
	\label{stdfactexamples}
	3241=(3)(24)(1),\hspace{0.2cm} 54321=(5)(4)(3)(2)(1),\hspace{0.2cm} 597634218 = (5976)(34)(2)(18),\hspace{0.2cm} 1\tau = (1\tau),
\end{equation}
where $\tau$ is any word in $S(\{2,\dotsc,n\})$. The coefficients $C_w$ are rational numbers given by
\begin{equation}
	C_w = \frac{(-1)^{d_w}}{|w|\binom{|w|-1}{d_w}}
\end{equation}
for any finite word $w$, whose letters $w_i$ are distinct positive integers up to $n$, where $|w|$ denotes the length of $w$, and $d_w$ is the descent number of $w$ defined by
\begin{equation}
	d_w = \#\{1\leq i\leq |w|-1~|~ w_i>w_{i+1}\}.
\end{equation}
The descent numbers of the first three examples in eq.~\eqref{stdfactexamples} are
\begin{equation}
	d_{3241}=2,\hspace{0.3cm} d_{54321}=4,\hspace{0.3cm} 
	d_{597634218}=5.
\end{equation}

It turns out that the terms in the decomposition~\eqref{trdecomp} cannot be linearly related to each other via the Jacobi identity. Recall that the set of linearly independent permutations of a string of structure constants $F^w$ is
\begin{equation}
	\{F^{w_1w_{\sig(2)}\cdots w_{\sig(n-1)}w_n}~|~\sig\in S(\{2,3,\dotsc,n-1\})\} .
\end{equation}
Acting with the Jacobi identity on such strings always takes one out of
the space of structures in \eqn{trdecomp}.  There is another set of relations
from the fact that the $d_R$'s are invariant tensors, but such relations all generate one term containing an $F^{n\cdots}$ which is also outside the space of structures.  In other words, the decomposition~\eqref{trdecomp} gives a basis of color factors.  Thus, we can use this decomposition to write $n$-point amplitudes of any gauge theory that have a color-decomposition of the form
\begin{equation}
	\sum_{\sig\in S_{n-1}} \tr_R(\sig n)A_n(\sig,n),
\end{equation}
with $A_n$ being color-ordered, in the color basis
\begin{equation}
	\label{casimirbasis}
	\{d_R^{nb_k\dots b_1}F^{\sig_kb_k}\cdots F^{\sig_1b_1}~|~\sig_k\cdots \sig_1=\sig\in S_{n-1}\}.
\end{equation}
Tree-level and one-loop-level amplitudes of $SU(N)$ gauge theory have a single-trace form when written in the trace basis and the DDM basis, respectively. So they can all be rewritten in the basis~\eqref{casimirbasis}.

Even though the set \eqref{casimirbasis} consists of color factors that are not related by the Jacobi identity, it well known that higher-order Casimirs can be related to lower-order ones. For example, in $SU(N)$, the Casimirs of order less than or equal to $N$ are independent, but the ones of order greater than $N$ depend on the independent ones. However, the set \eqref{casimirbasis} is still a color basis. Writing an $m$-th order Casimir as a unique sum over products of independent Casimirs still results in a distinct element of the set \eqref{casimirbasis}. As an example, consider $\tr(X^m)$ in $SU(3)$, for $m>3$. This Casimir can be written in terms of the two independent Casimirs $\tr(X^2)$ and $\tr(X^3)$ as
\begin{equation}
	\tr(X^m)=\sum_{\substack{p,q\in\mathbb{Z}_{\geq0}\\2p+3q=m}} a_{pq}\tr(X^2)^p\tr(X^3)^q
\end{equation}
for some numbers $a_{pq}$, with at least one being non-zero. Since no other Casimir (i.e.~$\tr(X^{m'})$ for $m'\neq m$) can have the same decomposition, the elements of \eqref{casimirbasis} remain independent.

It is possible to have a Lie algebra and/or one of its representations in which certain Casimirs vanish. The adjoint representation of $SU(N)$ and any representation of $SU(2)$, for example, have vanishing Casimirs of odd order. Clearly, in these cases, not all elements of \eqref{casimirbasis} are independent. However, those that do not vanish remain independent from each other in the sense of the previous paragraph. We will still refer to \eqref{casimirbasis} as a color basis, nonetheless.

In this color basis, the single-trace terms of an amplitude become
\begin{equation}
	\label{dFFbasis}
	\sum_{\sig\in S_{n-1}} \tr_R(\sig n)A_n(\sig,n)
	=\sum_{\sig_k\cdots\sig_1=\sig\in S_{n-1}}
	d_R^{nb_k\dots b_1}
	F^{\sig_kb_k}\cdots F^{\sig_1b_1}A^{dFF}_n(\sig),
\end{equation}
where $A^{dFF}_n$ are the subamplitudes in the new basis. They are given by
\begin{equation}
	\label{dFFsubamp}
	A_n^{dFF}(\sig=\sig_k\cdots\sig_1) 
	= \sum_{\tau\in S_{n-1}}C_{\tau^{-1}(\sig_1)}\cdots C_{\tau^{-1}(\sig_k)}A_n(\tau,n).
\end{equation}
Unlike the color-ordered subamplitudes $A_n$, the $A_n^{dFF}$ are no longer permutations of the arguments of a single subamplitude. The $A_n^{dFF}$ are best viewed as functions from $S_{n-1}$ to the $\mathbb{Q}$-vector space generated by the color-ordered subamplitudes $A_n(\tau,n)$.

\section{Relations from anomaly cancellation via matter}
\label{sec:matterrelns}
This section recapitulates section 3.2 of ref.~\cite{Dixon:2024mzh}, and it is included for a self-contained exposition. It does, however, include more details on the double-trace relations previously omitted from that work.

According to ref. \cite{Costello:2023vyy}, including Weyl fermions in the representation 
\begin{equation}
	\label{R0}
	R_0=8F\oplus8\bar{F}\oplus\asym\oplus\asyma
\end{equation}
will nullify the one-loop all-plus gluon amplitude. Here, $\asym$ denotes the antisymmetric tensor representation. This choice of representation satisfies the anomaly cancellation requirement that the quartic Casimir of a representation $R$ be equal to that of the adjoint representation
\begin{equation}
  \tr_R(X^4)=\tr_G(X^4).
\label{eq:anomcancelR}  
\end{equation}
An obvious choice for this would be $R=G$, but then the cancellation follows trivially from the supersymmetry Ward identity (SWI)
\begin{equation}
	\label{SWI}
	A_n^{[1/2]}(1,2,\dotsc,n) = -A_n^{[1]}(1,2,\dotsc,n),
\end{equation}
which holds for both the all-plus-helicity and one-minus-helicity configurations.

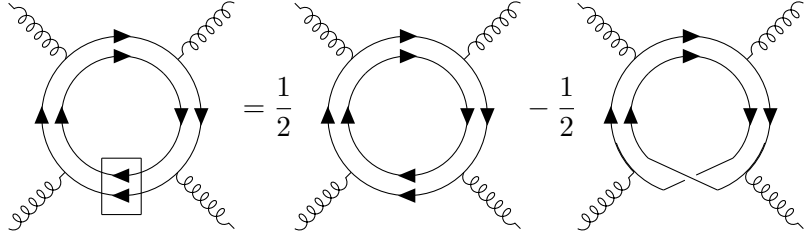
\begin{figure}
	\centering
	\begin{equation}
		\nonumber
		\begin{tikzpicture}[baseline={(0,0)},scale=0.75]
			\begin{feynman}
				\vertex (g1) at (0,-2);
				\vertex (g2) at (0,2);
				\vertex (g3) at (4,2);
				\vertex (g4) at (4,-2);
				\vertex (v1) at (1,-1);
				\vertex (v2) at (1,1);
				\vertex (v3) at (3,1);
				\vertex (v4) at (3,-1);
				\vertex (w1) at (1.25,-0.75);
				\vertex (w2) at (1.25,0.75);
				\vertex (w3) at (2.75,0.75);
				\vertex (w4) at (2.75,-0.75);
				\vertex (b1) at (1.65,-1.75);
				\vertex (b2) at (1.65,-0.75);
				\vertex (b3) at (2.35,-0.75);
				\vertex (b4) at (2.35,-1.75);
				\diagram*{
					(g1) -- [gluon] (v1),
					(v2) -- [gluon] (g2),
					(g3) -- [gluon] (v3),
					(v4) -- [gluon] (g4),
					(v1) -- [quarter left, fermion] (v2) 
					-- [quarter left, fermion] (v3) 
					-- [quarter left, fermion] (v4) 
					-- [quarter left, fermion] (v1),
					(w1) -- [quarter left, fermion] (w2) 
					-- [quarter left, fermion] (w3) 
					-- [quarter left, fermion] (w4) 
					-- [quarter left, fermion] (w1),
					(b1) -- (b2) -- (b3) -- (b4) -- (b1),
				};
			\end{feynman}
		\end{tikzpicture}
		=
		\frac{1}{2}
		\begin{tikzpicture}[baseline={(0,0)},scale=0.75]
			\begin{feynman}
				\vertex (g1) at (0,-2);
				\vertex (g2) at (0,2);
				\vertex (g3) at (4,2);
				\vertex (g4) at (4,-2);
				\vertex (v1) at (1,-1);
				\vertex (v2) at (1,1);
				\vertex (v3) at (3,1);
				\vertex (v4) at (3,-1);
				\vertex (w1) at (1.25,-0.75);
				\vertex (w2) at (1.25,0.75);
				\vertex (w3) at (2.75,0.75);
				\vertex (w4) at (2.75,-0.75);
				\diagram*{
					(g1) -- [gluon] (v1),
					(v2) -- [gluon] (g2),
					(g3) -- [gluon] (v3),
					(v4) -- [gluon] (g4),
					(v1) -- [quarter left, fermion] (v2) 
					-- [quarter left, fermion] (v3) 
					-- [quarter left, fermion] (v4) 
					-- [quarter left, fermion] (v1),
					(w1) -- [quarter left, fermion] (w2) 
					-- [quarter left, fermion] (w3) 
					-- [quarter left, fermion] (w4) 
					-- [quarter left, fermion] (w1),
				};
			\end{feynman}
		\end{tikzpicture}
		- \frac{1}{2}
		\begin{tikzpicture}[baseline={(0,0)},scale=0.75]
			\begin{feynman}
				\vertex (g1) at (0,-2);
				\vertex (g2) at (0,2);
				\vertex (g3) at (4,2);
				\vertex (g4) at (4,-2);
				\vertex (v1) at (1,-1);
				\vertex (v2) at (1,1);
				\vertex (v3) at (3,1);
				\vertex (v4) at (3,-1);
				\vertex (w1) at (1.25,-0.75);
				\vertex (w2) at (1.25,0.75);
				\vertex (w3) at (2.75,0.75);
				\vertex (w4) at (2.75,-0.75);
				\vertex (b1) at (2+1.4*0.34202,-1.4*0.93969);
				\vertex (b2) at (2-1.4*0.34202,-1.4*0.93969);
				\vertex (b3) at (2.1,-1.049161396);
				\vertex (b4) at (1.9,-1.141211056);
				\diagram*{
					(g1) -- [gluon] (v1),
					(v2) -- [gluon] (g2),
					(g3) -- [gluon] (v3),
					(v4) -- [gluon] (g4),
					(v1) -- [quarter left, fermion] (v2) 
					-- [quarter left, fermion] (v3) 
					-- [quarter left, fermion] (v4),
					(w1) -- [quarter left, fermion] (w2) 
					-- [quarter left, fermion] (w3) 
					-- [quarter left, fermion] (w4),
					(b1) -- (w1),
					(w4) -- (b3),
					(b4) -- (b2),
				};
				\draw ([shift=(-20:1.4)] 2, 0) arc (-20:-70:1.4);
				\draw ([shift=(200:1.4)] 2, 0) arc (200:250:1.4);
			\end{feynman}
		\end{tikzpicture}
	\end{equation}
	\caption{Color diagram for the trace over the representation $\asym$ in terms of traces over the fundamental $F$.}
	\label{fig:antisym}
\end{figure}

We will see that the choice $R=R_0$ implies nontrivial vanishing via linear relations among the all-plus subamplitudes.  The reason $R=R_0$ obeys \eqn{eq:anomcancelR} is the Casimir relation
\begin{equation}
  \label{antisymX4}
  \tr_\asym(X^4) = (N-8) \tr_F(X^4)\, +\, 3\, \bigl[\tr_F(X^2)\bigr]^2 \,,
\end{equation}
as can be seen using fig.~\ref{fig:antisym}; the ``$-8$'' comes from $(-1/2) \times 2^4$ contributions from the second ``exchange'' term on the right-hand side of fig.~\ref{fig:antisym}.  In \eqn{eq:anomcancelR}, after including also the $\asyma$ contributions, the leading-$N$ and the double-trace terms in \eqn{antisymX4} match corresponding terms from $\tr_G(X^4)$, while the $8F$ contribution cancels the ``$-8$''.

\begin{figure}[h]
	\centering
	\begin{equation}
		\nonumber
		8
		\begin{tikzpicture}[baseline={(0,0)},scale=0.75]
			\begin{feynman}
				\vertex (g1) at (0,-2);
				\vertex (g2) at (0,2);
				\vertex (g3) at (4,2);
				\vertex (g4) at (4,-2);
				\vertex (v1) at (1,-1);
				\vertex (v2) at (1,1);
				\vertex (v3) at (3,1);
				\vertex (v4) at (3,-1);
				\diagram*{
					(g1) -- [gluon] (v1),
					(v2) -- [gluon] (g2),
					(g3) -- [gluon] (v3),
					(v4) -- [gluon] (g4),
					(v1) -- [quarter left, fermion] (v2) 
					-- [quarter left, fermion] (v3) 
					-- [quarter left, fermion] (v4) 
					-- [quarter left, fermion] (v1),
				};
			\end{feynman}
		\end{tikzpicture}
		+8
		\begin{tikzpicture}[baseline={(0,0)},scale=0.75]
			\begin{feynman}
				\vertex (g1) at (0,-2);
				\vertex (g2) at (0,2);
				\vertex (g3) at (4,2);
				\vertex (g4) at (4,-2);
				\vertex (v1) at (1,-1);
				\vertex (v2) at (1,1);
				\vertex (v3) at (3,1);
				\vertex (v4) at (3,-1);
				\diagram*{
					(g1) -- [gluon] (v1),
					(v2) -- [gluon] (g2),
					(g3) -- [gluon] (v3),
					(v4) -- [gluon] (g4),
					(v1) -- [quarter right, fermion] (v4) 
					-- [quarter right, fermion] (v3) 
					-- [quarter right, fermion] (v2) 
					-- [quarter right, fermion] (v1),
				};
			\end{feynman}
		\end{tikzpicture}
		+
		\begin{tikzpicture}[baseline={(0,0)},scale=0.75]
			\begin{feynman}
				\vertex (g1) at (0,-2);
				\vertex (g2) at (0,2);
				\vertex (g3) at (4,2);
				\vertex (g4) at (4,-2);
				\vertex (v1) at (1,-1);
				\vertex (v2) at (1,1);
				\vertex (v3) at (3,1);
				\vertex (v4) at (3,-1);
				\vertex (w1) at (1.25,-0.75);
				\vertex (w2) at (1.25,0.75);
				\vertex (w3) at (2.75,0.75);
				\vertex (w4) at (2.75,-0.75);
				\vertex (b1) at (1.65,-1.75);
				\vertex (b2) at (1.65,-0.75);
				\vertex (b3) at (2.35,-0.75);
				\vertex (b4) at (2.35,-1.75);
				\diagram*{
					(g1) -- [gluon] (v1),
					(v2) -- [gluon] (g2),
					(g3) -- [gluon] (v3),
					(v4) -- [gluon] (g4),
					(v1) -- [quarter left, fermion] (v2) 
					-- [quarter left, fermion] (v3) 
					-- [quarter left, fermion] (v4) 
					-- [quarter left, fermion] (v1),
					(w1) -- [quarter left, fermion] (w2) 
					-- [quarter left, fermion] (w3) 
					-- [quarter left, fermion] (w4) 
					-- [quarter left, fermion] (w1),
					(b1) -- (b2) -- (b3) -- (b4) -- (b1),
				};
			\end{feynman}
		\end{tikzpicture}
		+
		\begin{tikzpicture}[baseline={(0,0)},scale=0.75]
			\begin{feynman}
				\vertex (g1) at (0,-2);
				\vertex (g2) at (0,2);
				\vertex (g3) at (4,2);
				\vertex (g4) at (4,-2);
				\vertex (v1) at (1,-1);
				\vertex (v2) at (1,1);
				\vertex (v3) at (3,1);
				\vertex (v4) at (3,-1);
				\vertex (w1) at (1.25,-0.75);
				\vertex (w2) at (1.25,0.75);
				\vertex (w3) at (2.75,0.75);
				\vertex (w4) at (2.75,-0.75);
				\vertex (b1) at (1.65,-1.75);
				\vertex (b2) at (1.65,-0.75);
				\vertex (b3) at (2.35,-0.75);
				\vertex (b4) at (2.35,-1.75);
				\diagram*{
					(g1) -- [gluon] (v1),
					(v2) -- [gluon] (g2),
					(g3) -- [gluon] (v3),
					(v4) -- [gluon] (g4),
					(v1) -- [quarter right, fermion] (v4) 
					-- [quarter right, fermion] (v3) 
					-- [quarter right, fermion] (v2) 
					-- [quarter right, fermion] (v1),
					(w1) -- [quarter right, fermion] (w4) 
					-- [quarter right, fermion] (w3) 
					-- [quarter right, fermion] (w2) 
					-- [quarter right, fermion] (w1),
					(b1) -- (b2) -- (b3) -- (b4) -- (b1),
				};
			\end{feynman}
		\end{tikzpicture}
	\end{equation}
	\caption{The one-loop color diagram for matter in the representation $R_0=8F\oplus\bar{F}\oplus\asym\oplus\asyma$.}
	\label{fig:R1loop}
\end{figure}
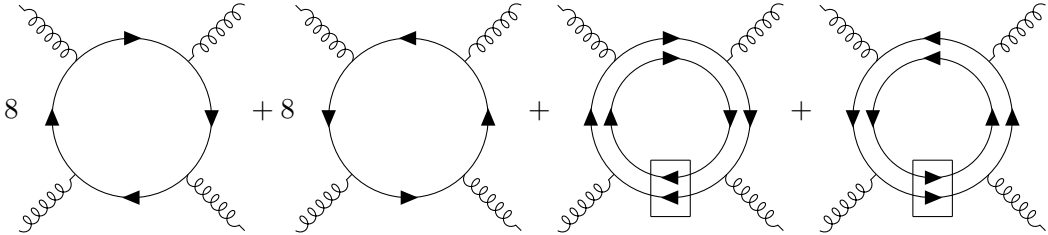

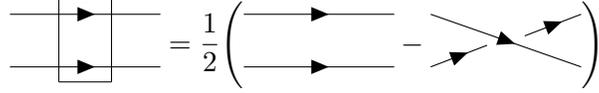
\begin{figure}
	\centering
	\begin{equation}
		\nonumber
		\begin{tikzpicture}[baseline={(0,0)},scale=1]
			\begin{feynman}
				\vertex (v1) at (1,-0.2);
				\vertex (v2) at (3,-0.2);
				\vertex (w1) at (1,0.5);
				\vertex (w2) at (3,0.5);
				\vertex (b1) at (1.65,-0.4);
				\vertex (b2) at (1.65,0.7);
				\vertex (b3) at (2.35,0.7);
				\vertex (b4) at (2.35,-0.4);
				\diagram*{
					(v1) -- [fermion] (v2),
					(w1) -- [fermion] (w2),
					(b1) -- (b2) -- (b3) -- (b4) -- (b1),
				};
			\end{feynman}
		\end{tikzpicture}
		=
		\frac{1}{2}
		\Bigg(
		\begin{tikzpicture}[baseline={(0,0)},scale=1]
			\begin{feynman}
				\vertex (v1) at (1,-0.2);
				\vertex (v2) at (3,-0.2);
				\vertex (w1) at (1,0.5);
				\vertex (w2) at (3,0.5);
				\diagram*{
					(v1) -- [fermion] (v2),
					(w1) -- [fermion] (w2),
				};
			\end{feynman}
		\end{tikzpicture}
		-
		\begin{tikzpicture}[baseline={(0,0)},scale=1]
			\begin{feynman}
				\vertex (v1) at (1,-0.2);
				\vertex (v2) at (3,-0.2);
				\vertex (w1) at (1,0.5);
				\vertex (w2) at (3,0.5);
				\vertex (v1p) at (1.75,0.0875);
				\vertex (w2p) at (2.25,0.2125);
				\diagram*{
					(v1) -- [fermion] (v1p),
					(w2p) -- [fermion] (w2),
					(w1) -- [fermion] (v2),
				};
			\end{feynman}
		\end{tikzpicture}
		\Bigg)
	\end{equation}
	\caption{Graphical representation of the anti-symmetric tensor product of the fundamental representation in terms of two fundamental lines.}
	\label{fig:asym}
\end{figure}

In order to find closed-form expressions for these relations, we need to decompose the trace over generators in the $R_0$ representation in \eqn{R0} as traces over the fundamental representation. Decomposing the direct sums appearing in \eqn{R0} gives the trace over $R_0$ as
\begin{equation}
	\label{Rtrace}
	\tr_{R_0}(t^{a_1}\cdots t^{a_n})
	=~ 8\tr_F(1\cdots n) + 8(-1)^n\tr_F(n\cdots 1)
	+ \tr_{\asym}(1\cdots n) + \tr_{\asyma}(1\cdots n),
\end{equation}
where we used~\eqn{fundtracerefid}. The color diagram corresponding to $\tr_{R_0}$ is shown in fig.~\ref{fig:R1loop}. The rectangle covering the lines appearing in the diagrams denotes anti-symmetrization of those lines, as depicted in fig.~\ref{fig:asym}. The trace over the anti-symmetric tensor representation decomposes as traces over the fundamental by
\begin{align}
	\label{asymtrace}
\begin{split}
	\tr_{\asym}(1\cdots n) &= \frac{1}{2}\sum_{I\subset(1,\dotsc,n)}[\tr(I)\tr(I^c)-\tr(I\cdot I^c)]
	\\
	&= N\tr(1\cdots n) 
	+ \frac{1}{2}\sum_{\emptyset\neq I\subsetneq(1,\dotsc,n)}\tr(I)\tr(I^c)
	-\frac{1}{2}\sum_{I\subset(1,\dotsc,n)}\tr(I\cdot I^c),
\end{split}
\end{align}
where $I\cdot I^c$ means to concatenate the lists. This can be understood diagramatically with the double-line notation as depicted in fig.~\ref{fig:antisym}.

Plugging the decomposition \eqref{adjtoF} of the adjoint pure-gluon contribution with the $R_0$ matter contribution \eqref{Rtrace} and \eqref{asymtrace} into the DDM color decomposition \eqref{DDMwithRmatter} yields
\begin{align}
	\label{trGR}
	\begin{split}
		\tr_G(1\cdots n)&A_n^{[1]}(1,\dotsc,n)
		+ \tr_R(1\cdots n)A_n^{[1/2]}(1,\dotsc,n)
		\\
		=&~ -8\tr(1\cdots n)A_n^{[1]}(1,\dotsc,n) - 8\tr(n\cdots 1) A_n^{[1]}(n,\dotsc,1)
		\\
		&+\frac{1}{2}\sum_{I\subset(1,\dotsc,n)}
		\tr(I\cdot I^c)A_n^{[1]}(1,\dotsc,n) 
		+ \tr((I\cdot I^c)^T)A_n^{[1]}(n,\dotsc,1)
		\\
		&+\frac{1}{2}\sum_{\emptyset\neq I\subsetneq(1,\dotsc,n)}
		\big[
		2\tr(I)\tr((I^c)^T)-\tr(I)\tr(I^c)-\tr(I^T)\tr((I^c)^T)
		\big]A_n^{[1]}(1,\dotsc,n),
	\end{split}
\end{align}
where we have used the SWI \eqref{SWI} and the reflection identity \eqref{refid} obeyed by the subamplitude. The full amplitude is then given by the sum over all permutations on $n$ letters modulo permutations related by cycles and reflections. 

We define the subamplitude $A_n^{R_0}(1,\dotsc,n)$ to be the kinematic factor multiplying the single-trace color factor $\tr(1,\dotsc,n)$ after performing the sum of eq.~\eqref{trGR} over the appropriate permutations. It is given by
\begin{equation}
\label{AGR}
A_n^{R_0}(1,\dotsc,n) = - 8 A_n^{[1]}(1,\dotsc,n)
+ \sum_{k=1}^n\sum_{\sigma\,\in\,\alpha_k\shuffle\beta_k}A_n^{[1]}(1,\sigma),
\end{equation}
where $\alpha_k=(2,\dotsc,k)$ and $\beta_k=(k+1,\dotsc,n)$.
The first term comes from the trace over eight copies of the fundamental.
The remaining terms come from the exchange term in the trace over the
antisymmetric tensor representation,
\begin{equation}
	\label{trproj}
	\frac{1}{2}\tr_{F\otimes F}(1\cdots nP)
	= \frac{1}{2}\sum_{I\subset(1,\dotsc,n)}\tr(I\cdot I^c),
\end{equation}
where $P$ is the permutation operator that exchanges the two $F$
representations. In particular, the sum over $k$ appears since the list
$(1,\dotsc,k)=(1,\alpha_k)$ appears in the sum in eq.~\eqref{trproj}
for all $1\leq k\leq n$.  See ref.~\cite{Dixon:2024mzh} for a proof that the subamplitude $A_n^{R_0}(1,\dotsc,n)$ is given by eq.~\eqref{AGR}.

Now we turn to the double-trace terms in \eqn{trGR}.
The subamplitudes $A_{n;c}^{R_0}$ accompanying the double-trace terms
$\tr(1\cdots (c-1))\tr(c\cdots n)$ in \eqn{trGR} are given simply by
\begin{align}
	\label{AGRdbtr}
	\begin{split}
		A_{n;c}^{R_0}(\alpha,\beta) &= A_{n;c}(\alpha,\beta) - (-1)^{|\beta|}A_{n;c}(\alpha,\beta^T)
		\\
		&= A_{n;c}(\alpha,\beta) - (-1)^{|\alpha|}A_{n;c}(\alpha^T,\beta)
	\end{split},
\end{align}
where $\alpha=[1,2,\dotsc,c-1]$ and $\beta=[c,c+1,\dotsc,n]$, and the subamplitudes $A_{n;c}$ are given by eq.~\eqref{Anc}. This follows from the standard arguments for reversing the order of double-trace terms like those appearing in eq.~\eqref{trGR}~\cite{Bern:1994zx, DelDuca:1999rs}. Note that the double-trace terms coming from the conjugate of the antisymmetric tensor representation naturally appear in $A_{n;c}^{R_0}$, but they are related to the non-conjugated one via
\begin{equation}
	(-1)^{|\alpha|+|\beta|}\sum_{\sig\in\alpha^T\shuffle\beta^T}A_n^{[1]}(\sig)
	=\sum_{\sig\in\alpha^T\shuffle\beta^T}A_n^{[1]}(\sig^T)
	=\sum_{\sig\in\alpha\shuffle\beta}A_n^{[1]}(\sig),
\end{equation}
where the first equality uses the reflection identity \eqref{refid}. The second line of eq.~\eqref{AGRdbtr} follows from applying the reflection identity \eqref{refid} on the first line.

Since the full all-plus amplitude vanishes for the fermion representation $R_0$ and since the traces over the generators are linearly independent in $SU(N)$ (up to dihedral symmetries), the right-hand sides of eqs.~\eqref{AGR} and \eqref{AGRdbtr} must also vanish. Thus, we have the all-plus single- and double-trace relations,
\begin{align}
	\label{singletrreln}
	0&=-8A_n^{[1]}(1,\dotsc,n) + \sum_{k=1}^n\sum_{\sigma\in\alpha_k\shuffle\beta_k}A_n^{[1]}(1,\sigma),
	\\
	\label{dbtrreln}
	0&=A_{n;c}(\alpha,\beta) - (-1)^{|\alpha|}A_{n;c}(\alpha^T,\beta)
	\hspace{0.5cm}~\text{for all}~2\leq c\leq n.
\end{align}
Remarkably, the first set of relations are exactly the same\footnote{
	The boundary terms $k=1$ and $k=n$ each just give $A_n^{[1]}(1,\dotsc,n)$. Removing them from the sum over $k$ converts the ``8'' to a ``6'' and puts \eqn{singletrreln} into the form in ref.~\cite{Bjerrum-Bohr:2011jrh}. }
all-plus relations conjectured previously~\cite{Bjerrum-Bohr:2011jrh}.
Ref.~\cite{Bjerrum-Bohr:2011jrh} based their formula on a decomposition of the all-plus subamplitudes into kinematic diagrams containing a single totally symmetric quartic vertex with the remaining vertices being all cubic and totally antisymmetric.  As we will see, this approach is closely related to the symmetrized-trace decomposition in \eqn{dFFbasis}.

The all-plus $n$-point one-loop color-ordered subamplitude is \cite{Bern:1993qk,Mahlon:1993si}
\begin{equation}
	\label{1loopsubamp}
	A_n^{[1]}(1,2,\dotsc,n) = -\frac{i}{48\pi^2}
	\frac{1}{\spdenom{n}}
	\sum_{1\leq i_1<i_2<i_3<i_4\leq n}\spa{i_1}.{i_2}\spb{i_2}.{i_3}\spa{i_3}.{i_4}\spb{i_4}.{i_1}.
\end{equation}
For general $n$, the number of terms appearing in the sum over $k$ in eq.~\eqref{singletrreln} is
\begin{equation}
	\sum_{k=1}^{n}\binom{n-1}{k-1}=2^{n-1},
\end{equation}
counting multiplicities.

For $n=4$, it is easy to verify \eqn{singletrreln},
because all $2^3=8$ terms in the sum over $k$ are equal,
thanks to the total symmetry of the four-point subamplitude,
\begin{equation}
	\label{4ptamp}
	A_4^{[1]}(1,2,3,4)=-\frac{i}{48\pi^2}
	\frac{\spb{2}.{3}\spb{4}.{1}}{\spa{2}.{3}\spa{4}.{1}},
\end{equation}
which follows from momentum conservation and the Schouten identity.
They cancel the remaining term.  Notice that if we take the relations as given, then we recover the fact that $A_4^{[1]}(1,2,3,4)$ is totally symmetric in its arguments. 

For $n>4$, eq.~\eqref{singletrreln} is not so easily verified from the explicit formula \eqref{1loopsubamp}. We have checked that it holds for $n\leq 11$ by replacing all spinor brackets with $3n-10$ independent momentum invariants, using a momentum-twistor parametrization~\cite{Badger:2013gxa,Hodges:2009hk}, which we describe in appendix~\ref{sec:twistorparam}.

The double-trace relations \eqref{dbtrreln} contain both previously known and unknown ones. The case $c=2$ is trivial because $A_{n;2} = 0$ by photon decoupling for an adjoint in the loop.  For $c=3$ and $n$ arbitrary, the sets $\alpha\shuffle\beta^T$ and $\alpha\shuffle\beta$ are in one-to-one correspondence via the map $\sig\mapsto\sig^T$, and the fact that $\alpha$ is equivalent to $\alpha^T$ for $|\alpha|=2$; that is, cyclic ordering is meaningless for a two-element set.  The relation then follows from the reflection identity \eqref{refid}.

For $c=4$, \eqn{dbtrreln} are equivalent to the three-photon-vanishing relations first observed in ref.~\cite{Bern:1993qk}. To see this, note that converting three gluons into photons amounts to summing over all possible insertions of the three gluons, while $A_{n;3}(\alpha,\beta)$ only contains half of these permutations, where the three elements of $\alpha$ are in a particular cyclic order.  The other half is given by reversing the cyclic order of $\alpha$, i.e.~by $A_{n;3}(\alpha^T,\beta)$.  The sum is the three-photon amplitude and it vanishes by \eqn{dbtrreln}.

For $c\geq5$, the relations are novel.  For example, the $c=5$ relations are distinct from four-photon vanishing (which follows from three-photon vanishing) because they sum over far fewer permutations (only two out of six) and with the opposite sign. They are not so readily verified. We have checked that they hold for $n\leq 11$ by using the momentum-twistor parametrization \eqref{twistparam}.

Solving these relations determines the number of linearly independent subamplitudes for small values of $n$. That is, for a given $n$, we solve all non-trivial permutations of eqs.~\eqref{singletrreln} and \eqref{dbtrreln} to determine the number of unconstrained subamplitudes implied by these relations. We can compare these numbers to the true number of linearly independent subamplitudes over the rational numbers, which can be determined analytically using a momentum-twistor parametrization. To get the true number, we consider the linear combination
\begin{equation}
	\label{lincom}
	0=\sum_{\sig\in S_n/R\mathbb{Z}_n}c_\sig A_n^{[1]}(\sig)
\end{equation}
where $c_\sig$ are arbitrary rational coefficients. The momentum-twistor parametrization expresses the subamplitudes as ratios of polynomials in the $3n-10$ independent momentum-twistor variables $x_i$. Since the monomials generated by the $x_i$ form a basis, eq.~\eqref{lincom} gives linear constraints on the $c_\sig$, once the sum is written as a single ratio of polynomials, where all the $c_\sig$ appear in one polynomial factor in the numerator. The true number of linearly independent subamplitudes is then the number of $c_\sig$ that are left unconstrained, after solving eq.~\eqref{lincom} for each monomial in the $x_i$.

We find that the single-trace relations \eqref{singletrreln} by themselves give the true number of linearly independent subamplitudes, whereas the double-trace relations eq.~\eqref{dbtrreln} are not as constraining. The various numbers of the unconstrained subamplitudes determined by these relations are given in table \ref{tab:allplustable}. In section~\ref{sec:axionrelns} we will argue that the total number of linearly independent $n$-point subamplitudes is given by an unsigned Stirling number of the first kind, $c(n-1,3)$. The value of $c(m,k)$ is the number of ways to partition a set of $m$ elements into $k$ distinct, cyclicly ordered sublists. For example, given the set ${1,2,3,4}$, the elements are placed into three sublists in the following six ways
\begin{equation}
	(12)(3)(4),~~(13)(2)(4),~~(14)(2)(3),~~(1)(23)(4),~~(1)(24)(3),~~(1)(2)(34),
\end{equation}
which means that $c(4,3)=6$. For the reader familiar with cycle notation for permutations, $c(m,k)$ is the number of permutations on $m$ letters with $k$ disjoint cycles. We provide some more background on the unsigned Stirling numbers in appendix~\ref{sec:factlength}.

\begin{table}[tbp]
	\centering
	\begin{tabular}{|c|c|c|c|c|c|}
		\hline
		$n$ & 4 & 5 & 6 & 7 & 8 \\
		\hline 
		$\# S_n/R\mathbb{Z}_n$ & 3 & 12 & 60 & 360 & 2520
                \\
		$c=4$ & 3 & 12 & 59 & 345 & 2344
		\\
		$c=5$ & 3 & 12 & 60 & 360 & 2429
		\\
		\text{single-trace} & 1 & 6 & 35 & 225 & 1624
		\\
		LI & 1 & 6 & 35 & 225 & 1624
		\\
		$c(n-1,3)$ & 1 & 6 & 35 & 225 & 1624
		\\
		\hline
	\end{tabular}
	\caption{The number of all-plus YM $n$-point subamplitudes $A_n^{[1]}(1,2,\ldots,n)$ that are linearly independent after applying various constraints.  The first line uses cyclic invariance and the reflection identity to get to $(n-1)!/2$ independent subamplitudes. Imposing the $c=4$ part of the double-trace relations \eqref{dbtrreln} only has an impact for $n\geq6$.  Similarly, imposing only the $c=5$ part has even less impact, and starts at $n=8$. Also, at $n=8$, the $c=5$ constraints are strictly weaker than the $c=4$ constraints. In contrast, the set of single-trace relations \eqref{singletrreln} alone reduce the number of linearly independent subamplitudes to that given on the fourth line. ``LI'' refers to the number of true linearly independent subamplitudes, as determined analytically using a momentum-twistor representation. It agrees with the previous line and with the unsigned Stirling number of the first kind, $c(n-1,3)$.}
\label{tab:allplustable}
\end{table}

\subsection{One-minus relations}

Equations~(\ref{AGR}) and (\ref{AGRdbtr}) hold for both of the one-loop amplitudes we consider in this paper, all-plus and one-minus.  In the all-plus case, $A_n^{R_0}$ vanishes.  As we will see in a second, $A_n^{R_0}$ is non-vanishing in the one-minus case. In both cases, the double-trace coefficients $A_{n;c}^{R_0}(\alpha,\beta)$ will vanish.

The one-loop amplitude with a single negative helicity in a theory with matter in the representation $R_0$ was also computed in ref.~\cite{Costello:2023vyy}, and it is given by
\begin{align}
	\label{1mtr}
	A_n^{R_0}(1^-,2^+,\dotsc,n^+)
	&=\frac{i}{48\pi^2}\times \frac{6}{\spdenom{n}} \times
	\sum_{2\leq i<j\leq n} \frac{\spb{i}.{j}}{\spa{i}.{j}}
	\, {\spa{1}.{i}}^2{\spa{1}.{j}}^2 \,,
	\\
	\label{1mdbtr}
	A_{n,c}^{R_0}(\alpha,\beta) &= 0.
\end{align}
Unlike the all-plus amplitude, the one-minus amplitude for the representation $R_0$ is non-vanishing, but it has a nice property: according to eq.~\eqref{1mdbtr}, it only has single-trace contributions in the trace basis. In other words, all double-trace terms vanish, so eq.~\eqref{dbtrreln} also holds in the one-minus case. The vanishing of the double-trace structure follows from the recursive construction used in ref.~\cite{Costello:2023vyy}, and its absence for the two- and three-point correlators since $\tr(t^a)=0$ in $SU(N)$.  The double-trace vanishing is also consistent with the fact that the last factor in \eqn{1mtr} is independent of the ordering of legs $2,3,\ldots,n$. So the dependence of $A_n^{R_0}(1^-,\sigma)$ on the ordering $\sigma$ is equivalent to that of the MHV tree (Parke-Taylor~\cite{Parke:1986gb,Mangano:1987xk}) amplitude.  Hence $A_n^{R_0}$ obeys all the Kleiss-Kuijf relations~\cite{Kleiss:1988ne} that $n$-gluon tree amplitudes obey, which leads to \eqn{1mdbtr}.  This fact also means that there are only $(n-2)!$ linearly independent permutations of $A_n^{R_0}$.

It was previously conjectured in ref.~\cite{Bjerrum-Bohr:2011jrh} that relations among the one-minus subamplitudes existed. They observed the following for low-multiplicity amplitudes: amplitude relations which are valid for both tree-level amplitudes and all-plus one-loop amplitudes are automatically also satisfied for the one-minus one-loop amplitudes.  We can see this property explicitly from the fact that the dependence of the ``inhomogeneous term'' $A_n^{R_0}(1^-,2^+,\ldots,n^+)$ in \eqn{AGR} on the ordering of $(2,3,\ldots,n)$ is equivalent to that of an MHV tree amplitude, so it vanishes after applying any tree-level (Kleiss-Kuijf) sum over permutations, returning us to the ``homogeneous'' all-plus types of relations.  This principle enabled the authors to predict the correct number of linearly independent ones, since this number would have to be the sum of the number of linearly independent tree-level amplitudes, $(n-2)!$, and all-plus one-loop amplitudes, $c(n-1,3)$. This principle also allowed them to determine that three-photon vanishing must hold as well for the one-minus amplitudes. However, the explicit relations~(\ref{AGR}) [with \eqref{1mtr}], and \eqref{1mdbtr} for $c>4$, were not known at the time.

The one-minus amplitude in QCD was computed first by Mahlon~\cite{Mahlon:1993si}.
We use the form given in ref.~\cite{Bern:2005ji}:
\begin{equation}
A_n^{[1]}(1^-,2^+,3^+,\ldots,n^+) =
{i\over48\pi^2} 
  {T_1 + T_2 \over \spa{1}.{2}\spa{2}.{3}\cdots \spa{n}.{1} } \,,
\label{oneminusalln}
\end{equation}
where
\begin{eqnarray}
 T_1 &=& \sum_{l=2}^{n-1} 
  { \spa{1}.{l} \spa{1,}.{l+1} 
    \sandmp1.{\Ksl_{l,l+1} \Ksl_{(l+1)\cdots n}}.1
   \over \spa{l,}.{l+1} } \,,
\label{oneminusallnT1} \\
 T_2 &=& \sum_{l=3}^{n-2} \sum_{p=l+1}^{n-1}
 { \spa{l-1,}.{l}
   \over \sandmp1.{\Ksl_{(p+1)\cdots n} \Ksl_{l\cdots p}}.{l-1}
         \sandmp1.{\Ksl_{(p+1)\cdots n} \Ksl_{l\cdots p}}.{l} }    
\nonumber \\
&& \hskip15mm\times 
 { \spa{p,}.{p+1}
   \over \sandmp1.{\Ksl_{2\cdots (l-1)} \Ksl_{l\cdots p}}.{p}
         \sandmp1.{\Ksl_{2\cdots (l-1)} \Ksl_{l\cdots p}}.{p+1} }
\nonumber \\
&& \hskip15mm\times
   {\sandmp1.{\Ksl_{l\cdots p} \Ksl_{(p+1)\cdots n}}.1}^3
\nonumber \\
&& \hskip15mm\times
   { \sandmp1.{\Ksl_{2\cdots (l-1)} [ {\cal F}(l,p) ]^2  \Ksl_{(p+1)\cdots n}}.1
    \over s_{l\cdots p} }
 \,,
\label{oneminusallnT2}
\end{eqnarray}
Here $s_{l\cdots p} = (k_l + k_{l+1} + \cdots + k_p)^2$ is a multi-particle Lorentz invariant, $\Ksl_A \equiv \sigma^\mu \sum_{a\in A} k_a^\mu$ is a sum of consecutive momenta written as a spinor matrix, so that
\begin{equation}
\spaa{i}.{\Ksl_A}.{\Ksl_B}.{j}
= \sum_{a\in A} \sum_{b\in B} \spa{i}.{a} \spb{a}.{b} \spa{b}.{j} \,,
\label{longerstrings}
\end{equation}
and
\begin{equation}
{\cal F}(l,p) = \sum_{i=l}^{p-1} \sum_{m=i+1}^{p} \ksl_i \ksl_m \,.
\label{Flpdef}
\end{equation}

Let us verify that the single-trace kinematic factor is given by eq.~\eqref{1mtr} for $n=4$. That is, we will check that the linear combination in eq.~\eqref{AGR} reduces to eq.~\eqref{1mtr} when eq.~\eqref{oneminusalln} is used for $n=4$.
After using momentum conservation on the expression \eqref{oneminusalln},
the 4-point one-minus subamplitude reduces to 
\begin{equation}
	\label{4pt1minus}
	A_4^{[1]}(1^-,2^+,3^+,4^+) = -\rho_{-}\frac{u^2}{st} \,,
\end{equation}
where
\begin{equation}
  \rho_{-}=\frac{i}{48\pi^2}
  \frac{\spa{1}.{2}\spa{1}.{4}\spb{2}.{4}}{\spa{2}.{3}\spa{3}.{4}\spa{2}.{4}}
\end{equation}
is a helicity-dependent overall phase that is totally symmetric in the arguments 2, 3, and 4. The kinematic variables $s$, $t$, and $u$ are the standard 4-point Mandelstam variables. After applying cyclic and reflection identities,
the single-trace kinematic factor \eqref{AGR} reduces to
\begin{equation}
A_4^{R_0}(1,2,3,4) = -4A_4^{[1]}(1,2,3,4) + 2A_4^{[1]}(1,3,2,4) + 2A_4^{[1]}(1,2,4,3).
\label{oneminus4ptsum}
\end{equation}
Inserting eq.~\eqref{4pt1minus} into the right-hand side of \eqn{oneminus4ptsum} gives
\begin{equation}
	A_4^{R_0}(1^-,2^+,3^+,4^+) = 2\rho_{-}
	\bigg(
	\frac{2u^2}{st} - \frac{s^2}{ut} - \frac{t^2}{su}
	\bigg)
	=6\rho_{-}\frac{s^2+st+t^2}{st}.
\end{equation}
The last equality follows after substituting $u=-s-t$ and simplifying the expression. The prediction \eqref{1mtr} in terms of $\rho_{-}$ and the Mandelstam variables is
\begin{equation}
	A_4^{R_0}(1^-,2^+,3^+,4^+)=6\rho_{-}\bigg(
	-\frac{u}{t} - 1 - \frac{u}{s}
	\bigg),
\end{equation}
where the first, second, and third terms within parentheses correspond to the terms in the sum appearing in eq.~\eqref{1mtr} with $(i,j)=(2,3)$, $(2,4)$, and $(3,4)$, respectively. Using $u=-s-t$, the two results agree.

Beyond $n=4$, we have checked that the relations \eqref{1mtr} and \eqref{1mdbtr} hold up to $n=11$, by using the above expressions for the one-minus amplitude and the momentum-twistor parametrization \eqref{twistparam}.

We have performed the same analysis on the one-minus amplitudes as we did for the all-plus ones, in order to determine the number of linearly independent subamplitudes. The results are shown in table \ref{tab:oneminustable}. Notice that the double-trace relations of eq.~\eqref{dbtrreln} completely determine all linear independent one-minus subamplitudes. Moreover, it is specifically the three-photon vanishing relations (when $c=4$ in the double-trace term) that appear to be the most constraining of the double-trace relations.

\begin{table}[tbp]
	\centering
	\begin{tabular}{|c|c|c|c|c|c|}
		\hline
		& $n=4$ & 5 & 6 & 7 & 8 \\
		\hline 
		$\# S_n/R\mathbb{Z}_n$ & 3 & 12 & 60 & 360 & 2520
		\\
		$c=4$ & 3 & 12 & 59 & 345 & 2344
		\\
		$c=5$ & 3 & 12 & 60 & 360 & 2429
		\\
		LI & 3 & 12 & 59 & 345 & 2344
		\\
		$c(n-1,3)+(n-2)!$ & 3 & 12 & 59 & 345 & 2344
		\\
		\hline
	\end{tabular}
	\caption{The number of linearly independent one-minus YM $n$-point subamplitudes after applying various constraints. The constraints imposed are the same as in the all-plus case in table~\ref{tab:allplustable}.  The main difference is that the inhomogeneous term $A_n^{R_0}$ in \eqn{AGR} is nonvanishing, and there are $(n-2)!$ such terms. Therefore the entries in the last two lines are larger than in table~\ref{tab:allplustable} by $(n-2)!$.}
\label{tab:oneminustable}
\end{table}

\section{Amplitudes with axion-exchange}
\label{sec:axion}
The first mechanism discussed for cancelling the sdYM anomaly in twistor space was to introduce a fourth-order ``axion'', i.e.~a scalar with a fourth-order kinetic term that couples to gluons via a three-point interaction and exhibits shift symmetry~\cite{Costello:2021bah,Costello:2022wso,Costello:2022upu}.  The anomaly cancellation resembles the Green-Schwarz mechanism in superstring theory~\cite{Green:1984sg}. The tree-level $n$-gluon amplitudes with a single internal axion propagator exactly cancel the one-loop pure-sdYM amplitudes for specific gauge groups that lack an independent quartic Casimir.  If there is an independent quartic Casimir, then a combination of the axion and Weyl fermions in specific representations $R$ can also be used to annihilate the all-plus amplitude \cite{Costello:2022upu,Costello:2022wso,Costello:2023vyy}.

The space-time action describing the axion is given by\footnote{The difference in the normalization of the coefficient of the coupling term from that used in refs.~\cite{Costello:2021bah,Costello:2022wso} arises from our choice of the normalization of the one-loop amplitude in eq.~\eqref{1loopsubamp}.}
\begin{equation}
	\label{axionaction}
	\int d^4x \biggl[
          -\frac{1}{2}(\square\rho)^2
          + \frac{\lam_{G,R}g^2}{4\pi\sqrt{3}}\rho~\tr(F\wedge F) \biggr],
\end{equation}
where $\square \equiv \partial_\mu\partial^\mu$, $\rho$ is the axion field and $F$ is the YM field strength 2-form, which is given in components by
\begin{equation}
	F_{\mu\nu}=\partial_\mu A_\nu - \partial_\nu A_\mu 
	- i\frac{g}{\sqrt{2}}[A_\mu,A_\nu] \,.
\end{equation}
When using a combination of the axion and Weyl fermions to cancel the anomaly, the Weyl fermions need to cancel the single-trace part of the anomaly.  Then the difference between the quartic Casimir in the adjoint and the one in the matter representation is proportional to the square of the fundamental quadratic Casimir, and the coupling constant $\lam_{G,R}$ required is determined by the constant of proportionality,
\begin{equation}
	\label{casaxplusmatter}
	\tr_G(X^4)-\tr_R(X^4)=\lam_{G,R}^2\, \bigl[ \tr_F(X^2)\bigr]^2 \,.
\end{equation}
For $SU(N)$ gauge theory, a combined axion-fermion solution for arbitrary $N$ is to include $n_f=N$ flavors of fermions in the fundamental representation (quarks), i.e.~$R=N\,(F\oplus \bar{F})$. The constant takes the value $\lam_{G,R}^2=6$ in this case.

If we wanted to only include the axion, and have no fermions, then the anomaly-cancellation condition becomes
\begin{equation}
	\label{casimiraxion}
	\tr_G(X^4)=\lam_G^2\tr_F(X^2)^2.
\end{equation}
Now the relation holds for gauge groups $SU(2)$, $SU(3)$, and the exceptional ones. The proportionality constant is given by \cite{Okubo:1978qe}
\begin{equation}
	\label{lamG}
	\lam_G^2=\frac{10(h^\vee)^2}{\dim G+2},
\end{equation}
where $h^\vee$ is the dual Coxeter number, equal to $N$ for $SU(N)$. The exact value of $\lam_G^2$ will not play a role in our analysis, except when we restrict ourselves to $SU(N)$ gauge theory with $N=2,3$.

\subsection{The $n$-point axion-exchange amplitude}
\label{sec:axionamps}

The tree-level axion-exchange amplitudes, denoted $\Ax_n$, are constructed fairly easily. The axion propagator splits any color-ordered amplitude into two tree-level color-ordered axion-to-gluon amplitudes, where the axion is off-shell. The color-ordered amplitude is then recovered as the product of these two off-shell axion-to-gluon amplitudes divided by the fourth-power of the momentum flowing through the axion propagator. Since the axion is colorless, the associated color factor is simply given by the product of the color factors associated to the two axion-to-gluon amplitudes, and hence it is purely double-trace. 

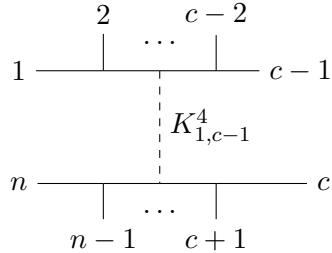
\begin{figure}[h]	
	\centering
	\begin{tikzpicture}[scale=0.75]
		\begin{feynman}
			\vertex (v1) at (-2.5,1) {\(1\)};
			\vertex (v2) at (-1,1);
			\vertex (v3) at (1,1);
			\vertex (v4) at (2.5,1) {\(c-1\)};
			\vertex (a1) at (0,1);
			\vertex (a2) at (0,-1);
			\vertex (v5) at (-2.5,-1) {\(n\)};
			\vertex (v6) at (-1,-1);
			\vertex (v7) at (1,-1);
			\vertex (v8) at (2.9,-1) {\(c\)};
			\vertex (g2) at (-1,2) {\(2\)};
			\vertex (g3) at (1,2) {\(c-2\)};
			\vertex (g6) at (-1,-2) {\(n-1\)};
			\vertex (g7) at (1,-2) {\(c+1\)};
			\vertex (vdot1) at (0,1.5) {\(\dots\)};
			\vertex (vdot2) at (0,-1.5) {\(\dots\)};
			\diagram*{
				(v1) -- (v2) -- (a1) -- (v3) -- (v4),
				(g2) -- (v2),
				(g3) -- (v3),
				(a1) -- [scalar, edge label=\(K_{1,c-1}^4\)] (a2),
				(v5) -- (v6) -- (a2) -- (v7) -- (v8),
				(g6) -- (v6),
				(g7) -- (v7),
				(a1) -- [opacity=0] (vdot1),
				(a2) -- [opacity=0] (vdot2),
			};
		\end{feynman}
	\end{tikzpicture}
	\caption{The double-comb color diagram $F^{1\dotsc (c-1)}F^{c\dotsc n}$ associated to the color-ordered amplitude $\ax_{n;c}(\alpha,\beta)$. The dotted line represents the colorless scalar connecting the two associated kinematic factors.}
	\label{fig:axion}
\end{figure}

The color factors associated to the partial amplitude $\ax_{n;c}(\alpha,\beta)$ for the contribution shown in Fig.~\ref{fig:axion} are given as the product of two color factors associated to $A_{c-1}^\rho(\alpha)$ and $A_{n-c+1}^\rho(\beta)$, where $c$ signifies that the axion splits the first $c-1$ gluons from the last $n-c+1$ gluons, and in the standard ordering $\alpha=[1,2,\dotsc,c-1]$ and $\beta=[c,c+1,\dotsc,n]$. In the DDM basis, the color factors are given by strings of $SU(N)$ structure constants $F^\alpha$ and $F^\beta$, respectively, or by traces over the fundamental $\tr(\alpha)$ and $\tr(\beta)$, respectively, in the trace basis. The full amplitude is then given by
\begin{equation}
	\label{fullaxamp}
	\Ax_n = g^n \lam_{G,R}^2\sum_{c=2}^{\floor{n/2}+1}\sum_{\sig\in S_n/S_{n;c}^{FF}}
			F^{\sig(1\dots (c-1))}F^{\sig(c\dotsc n)}
			\ax_{n;c}(\sig)
\end{equation}
in the DDM basis, or by 
\begin{equation}
	\label{axamptrbasis}
	\Ax_n = g^n\lam_{G,R}^2 \sum_{c=2}^{\floor{n/2}+1}
	\sum_{\sig\in S_n/S_{n;c}} \tr(\sig(1\cdots (c-1)))
	\tr(\sig(c\cdots n))\ax_{n;c}(\sig)
\end{equation}
in the trace basis. The set $S_{n;c}^{FF}$ contains all permutations that leave the double-comb structure invariant, and $S_{n;c}$ is the set of permutations that leave the double-trace structure invariant, as before.  The two color representations are equivalent because the axion-to-gluon tree amplitudes obey Kleiss-Kuijf relations.

The color-ordered axion-exchange subamplitude multiplying the double-trace color factor is given by
\begin{equation}
	\label{axampinit}
	\ax_{n;c}(\alpha,\beta) = A_{c-1}^\rho(\alpha)
	\frac{-i}{K_{1,c-1}^2K_{c,n}^2}
	A_{n-c+1}^\rho(\beta) \,.
\end{equation}
The momentum flowing through the axion propagator is $K_{1,c-1} = - K_{c,n}$, where $K_{i,j}=\sum_{m=i}^j k_m$ and $k_m$ is the momentum of external gluon $m$. We have used momentum conservation to write $K_{1,c-1}^4=K_{c,n}^4=K_{1,c-1}^2K_{c,n}^2$, which allows for the symmetric form shown in eq.~\eqref{axampinit}. The $A_n^\rho$ are the color-ordered axion-to-gluon amplitudes mentioned above; they were originally computed~\cite{Dixon:2004za,Berger:2006sh} as the amplitudes coming from an effective action that couples the Higgs boson (and a pseudoscalar partner) to two gluons after integrating out the top quark. With the normalization convention here, they are given by
\begin{equation}
	\label{axtoglu}
	A^\rho_n(1,\dotsc,n) = \frac{i}{4\pi\sqrt{3}}\frac{K_{1,n}^4}{\spa{1}.{2}\spa{2}.{3}\cdots\spa{n}.{1}}.
\end{equation}
So, eq.~\eqref{axampinit} is explicitly
\begin{equation}
	\label{axamp}
	\ax_{n;c}(1,\dotsc,n)=\frac{i}{48\pi^2}\frac{K_{1,c-1}^2}{\spa{1}.{2}\cdots \spa{c-1,}.{1}}
	\frac{K_{c,n}^2}{\spa{c,}.{c+1}\cdots \spa{n}.{c}}.
\end{equation}
For $n=4$ and $c=3$, eq.~\eqref{axamp} is
\begin{equation}
  \ax_{4;3}(1,2,3,4)
  = \frac{i}{48\pi^2}\frac{\spa{1}.{2}\spb{2}.{1}}{\spa{1}.{2}\spa{2}.{1}}
                     \frac{\spa{3}.{4}\spb{4}.{3}}{\spa{3}.{4}\spa{4}.{3}}
  = \frac{i}{48\pi^2}\frac{\spb{1}.{2}\spb{3}.{4}}{\spa{1}.{2}\spa{3}.{4}},
\end{equation}
which is equal and opposite to the four-point all-plus amplitude \eqref{4ptamp},
after permuting the arguments of this totally symmetric subamplitude.
Actually $\ax_{4;3}(1,2,3,4)$ should be compared not with $A_4^{[1]}$ but
with $A_{4;3}$ which is 6 times larger, according to \eqn{Anc}:
\begin{equation}
A_{4;3}(1,2,3,4) = - \frac{i}{48\pi^2}
\times 6 \times \frac{\spb{1}.{2}\spb{3}.{4}}{\spa{1}.{2}\spa{3}.{4}},
\end{equation}
The factor of 6 is accounted for, in the mixed axion-plus-$(n_f=N)$ case,
by the factor of $\lambda_{G,R}^2 = 6$ in \eqn{axamptrbasis}.

Similarly, one can compute the five-point double-trace contribution $A_{5;3}$
from \eqn{1loopsubamp} and the permutation sum~\eqref{Anc}:
\begin{equation}
A_{5;3}(1^+,2^+;3^+,4^+,5^+) = -\frac{i}{48\pi^2} \times 6
\times \frac{s_{12}}{\spa{1}.{2}\spa{2}.{1}} \,
       \frac{s_{12}}{\spa{3}.{4}\spa{4}.{5}\spa{5}.{3}} \,,
\label{A53}
\end{equation}       
which again cancels against \eqn{axampinit} after taking into account $\lambda_{G,R}^2 = 6$.

A comment should be made about the fact that the axion propagator comes with the opposite sign to that of a scalar propagator with a standard kinetic term. This minus sign follows directly from the minus sign in the kinetic term of the action \eqref{axionaction}. It is needed so that $\ax_{4;3}$ comes with the opposite sign of the four-point all-plus subamplitude $A_4^{[1]}$, while keeping the Lagrangian real-valued. The correct choice of sign to give the kinetic term seems arbitrary without the knowledge that the axion-exchange amplitude needs to cancel the all-plus one-loop one, but it is consistent with the fact that the propagator is the Green's function for the differential operator $\square^2$:
\begin{equation}
	\square_x^2\int\frac{d^4p}{(2\pi)^4}\frac{-i}{p^4}e^{ip\cdot(x-y)} = -i\delta(x-y),
\end{equation}
Compare this to a standard kinetic term $-\tfrac{1}{2}\rho\square\rho$, where the differential operator $\square$ satisfies
\begin{equation}
	\square_x\int\frac{d^4p}{(2\pi)^4}\frac{i}{p^2}e^{ip\cdot(x-y)} = -i\delta(x-y).
\end{equation}

\subsection{Axion and matter cancellation}
\label{sec:axplusmatter}

Adding both the axion and Weyl fermions renders the amplitudes of sdYM null, so long as the fermions live in a representation that satisfies eq.~\eqref{casaxplusmatter} \cite{Costello:2022upu,Costello:2022wso,Costello:2023vyy}. We will focus on $SU(N)$ gauge theory with $R=N\,(F\oplus \bar{F})$, for which $\lam_{G,R}^2=6$.

For this theory, the matter contribution eliminates the single-trace terms in the trace-basis color-decomposition in eq.~\eqref{tracebasis}, due to the supersymmetry Ward identity $A^{[1/2]}=-A^{[1]}$.  On the other hand, the axion contribution cancels the double-trace terms, leading to the relation
\begin{equation}
	\label{axequalsdbtr}
	A_{n;c}(\alpha,\beta) = -6\ax_{n;c}(\alpha,\beta),
\end{equation}
which is valid for all values of $c$.  We saw above how this relation is satisfied for $n=4$ and $n=5$ and $c=3$.

This relation simply tells us that the linear combination \eqref{Anc} describing the double-trace subamplitudes $A_{n;c}$ simplifies to the axion-exchange subamplitudes $\ax_{n;c}$ given in eq.~\eqref{axamp}. We will, however, leverage eq.~\eqref{axequalsdbtr} to help with revealing relations from cancelling the anomaly with only the axion.

\subsection{Relations from anomaly cancellation via the axion}
\label{sec:axionrelns}

According to refs.~\cite{Costello:2021bah,Costello:2022upu,Costello:2022wso}, the tree-level axion-exchange amplitude of eq.~\eqref{fullaxamp} cancels the all-plus one-loop pure YM amplitude $\mathcal{A}_n^{(1)}$ for gauge groups $SU(2)$ and $SU(3)$. In symbols,
\begin{equation}
	\mathcal{A}_n^{(1)}+\mathcal{A}_n^\text{ax}=0.
\end{equation}
In order to relate the single-trace structure of $\mathcal{A}_n^{(1)}$ to the double-trace structure of $\mathcal{A}_n^\text{ax}$, one needs to use the relation~\eqref{casimiraxion} between the quartic Casimir in the adjoint representation and the quadratic Casimir in the fundamental. In terms of traces of the generators, this relation is
\begin{equation}
	\label{casimirid}
	\tr_G(F^{(a}F^bF^cF^{d)})=\lam_G^2~\tr(T^{(a}T^b)\tr(T^cT^{d)})
	=\lam_G^2~\delta^{(ab}\delta^{cd)}.
\end{equation}
Since we need to apply a relation that involves symmetrized traces over the adjoint representation, we convert the all-plus one-loop amplitude from the DDM basis to the symmetrized-trace (Casimir) basis using eqs.~\eqref{dFFbasis} and \eqref{dFFsubamp}:
\begin{equation}
	g^n\sum_{\sig\in S_{n-1}} \tr_G(\sig,n)A_n^{[1]}(\sig,n)
	=g^n\sum_{\sig_k\cdots\sig_1=\sig\in S_{n-1}}
	d_G^{nb_k\dots b_1}
	F^{\sig_kb_k}\cdots F^{\sig_1b_1}A^{dFF}_n(\sig).
\end{equation}
The reflection identity $\tr_G(12\cdots n)=(-1)^n\tr_G(n\cdots21)$ implies that $d_G^{1\dots n}$ vanishes whenever $n$ is odd. So, the nontrivial contributions come from permutations with standard factorizations of odd length.

The case of most interest is when permutations have standard factorizations of length 3, as this corresponds to a symmetrized trace on four generators to which we can apply the identity~\eqref{casimirid}. These terms become
\begin{multline}
	d_G^{nb_1b_2b_3}F^{\sig_1b_1}F^{\sig_2b_2}F^{\sig_3b_3}
	\\
	=-\frac{\lam_G^2}{3}
	\Big((-1)^{|\sig_3|}F^{\sig_1n}F^{\sig_2\sig_3^T}
	    +(-1)^{|\sig_1|}F^{\sig_2n}F^{\sig_3\sig_1^T}
	    +(-1)^{|\sig_2|}F^{\sig_3n}F^{\sig_1\sig_2^T}\Big).
\end{multline}
From here, we can reorder the strings of $F$ matrices so that the smallest integer appears first and the largest appears last. Doing so converts the one-loop amplitude to the form
\begin{align}
\begin{split}
	\label{A1casbasis}
	\mathcal{A}_n^{(1)}
	=&~g^n\sum_{\substack{\sig_k\cdots\sig_1=\sig\in S_{n-1} \\ k\neq3}} d_G^{nb_k\dots b_1}F^{\sig_kb_k}\cdots F^{\sig_1b_1}A^{dFF}_n(\sig)
	\\
	&+ g^n\lam_G^2\sum_{c=2}^{\floor{n/2}+1}\sum_{\sig\in S_n/S_{n;c}^{FF}} F^{\sig(1)\dotsc\sig(c-1)}F^{\sig(c)\dotsc\sig(n)}
	A_{n;c}^{FF}(\sig),
\end{split}
\end{align}
where $A_{n;c}^{FF}$ is given by
\begin{equation}
	A_{n;c}^{FF}(\alpha,\beta) = \frac{1}{6}\sum_{(\sig,\tau)\in S_{c-1}/\Z_{c-1}\times S(c,\dotsc,n)/\Z_{n-c+1}}
	C_{\sig^{-1}\alpha}C_{\tau^{-1}\beta}A_{n;c}(\sig,\tau).
\end{equation}

The one-loop amplitude is now in a form of independent color-factors from which the relations implied by $\mathcal{A}_n^{(1)}+\mathcal{A}^\text{ax}_n=0$ are manifest. They are
\begin{align}
	\label{axdbtrrlens}
	0 &= A_{n;c}^{FF} + \ax_{n;c}(1,\dotsc,n),
	\\
	\label{dFFrelns}
	0 &= A_n^{dFF}(\sig)\hspace{1cm}
	\text{for all}~\sig_k\cdots\sig_1=\sig\in S_{n-1}~\text{with}~k~\text{odd and}~k\neq3.
\end{align}
The first equation relates the one-loop pure YM subamplitudes to the tree-level axion-exchange subamplitudes. However, from eq.~\eqref{axequalsdbtr}, we know that the axion subamplitudes are just the simplification of the all-plus double-trace terms. So, eq.~\eqref{axdbtrrlens} can be written as
\begin{equation}
	\label{AFFrelns}
	A^{FF}_{n;c} = \frac{1}{6}A_{n;c}(\alpha,\beta).
\end{equation}
The second equation \eqref{dFFrelns} also gives relations among the one-loop YM subamplitudes themselves. We have verified that both of these relations hold for $n\leq 8$ using the momentum-twistor parametrization \eqref{twistparam}. Interestingly though, eq.~\eqref{dFFrelns} holds for even values of $k$ too. However, this cannot be deduced from the decomposition \eqref{A1casbasis}, since the Casimirs $d_G^{nb_k\dotsc b_1}$ vanish for even values of $k$, by the reflection identity~\eqref{adjtracerefid}.

If we start with the amplitude in the trace basis \eqref{tracebasis} and convert the single-trace terms to the Casimir basis, using \eqn{dFFbasis} for $R=F$, then we get
\begin{align}
\begin{split}
	\mathcal{A}_n^{(1)}=&~g^nN\sum_{\sig_k\cdots\sig_1=\sig\in S_{n-1}}
	d_F^{nb_k\dots b_1}
	F^{\sig_kb_k}\cdots F^{\sig_1b_1}A^{dFF}_n(\sig)
	\\
	&+g^n\sum_{c=2}^{\lfloor n/2 \rfloor+1}\sum_{\sigma\in S_n/S_{n;c}}\tr(\sigma(1\dots (c-1)))\tr(\sigma(c\dots n))
	A_{n;c}(\sigma).
\end{split}
\end{align}
For $SU(2)$ and $SU(3)$, the quartic Casimir in the fundamental is proportional to the quadratic one with a factor of $1/2$, i.e.~$\tr(X^4) = \tfrac{1}{2}[\tr(X^2)]^2$. When we use this identity to move the $k=3$ terms from the first line to the second line, we get
\begin{align}
	\label{A1casfun}
\begin{split}
	\mathcal{A}_n^{(1)}=&~g^nN\sum_{\substack{\sig_k\cdots\sig_1=\sig\in S_{n-1} \\ k\neq3}}d_F^{nb_k\dots b_1}
	F^{\sig_kb_k}\cdots F^{\sig_1b_1}A^{dFF}_n(\sig)
	\\
	&+g^n\sum_{c=2}^{\lfloor n/2 \rfloor+1}\sum_{\sigma\in S_n/S_{n;c}}\tr(\sigma(1\dots (c-1)))\tr(\sigma(c\dots n))
	\bigg(A_{n;c}(\sigma)+NA_{n;c}^{FF}\bigg).
\end{split}
\end{align}
Note that the $A_{n;c}^{FF}$ term comes with a factor of $1/2$ from converting the quartic Casimir to the quadratic one, but this gets cancelled by a factor of 2 coming from converting the double-comb structure to a double-trace structure.

By use of eqs.~\eqref{axdbtrrlens} and \eqref{axequalsdbtr}, the double-trace term of eq.~\eqref{A1casfun} becomes
\begin{equation}
	(-6-N)A_{n;c}^\text{ax}=-\lam_G^2A_{n;c}^\text{ax} \,,
\end{equation}
where we used the fact that $\lam_G^2=N+6$ for $SU(2)$ and $SU(3)$. Adding the axion-exchange amplitude in the trace basis exactly cancels the double-trace term of eq.~\eqref{A1casbasis}, as expected. Thus, what remains must vanish
\begin{equation}
	0=\mathcal{A}_n^{(1)}+\mathcal{A}_n^\text{ax}=~g^nN\sum_{\substack{\sig_k\cdots\sig_1=\sig\in S_{n-1} \\ k\neq3}}d_F^{nb_k\dots b_1}
	F^{\sig_kb_k}\cdots F^{\sig_1b_1}A^{dFF}_n(\sig).
\end{equation}
In the fundamental representation of $SU(3)$, the Casimirs are generically non-vanishing, and, in particular, the odd-order ones do not vanish, unlike in the fundamental representation of $SU(2)$ or the adjoint representation of $SU(N)$. Thus, we can finally conclude that
\begin{equation}
	\label{dFFrelnsnew}
	0 = A_n^{dFF}(\sig)\hspace{1cm}
	\text{for all}~\sig_k\cdots\sig_1=\sig\in S_{n-1}~\text{for}~k\neq3,
\end{equation}
for $k$ even as well. Recall that the $A_n^{dFF}$ are expressed in terms of the single-trace coefficients $A_n$ in \eqn{dFFsubamp}.

The relations \eqref{dFFrelnsnew} appear to be the same relations obeyed by the $\zeta_2$ terms of tree-level open superstring (disk) amplitudes \cite{Mafra:2021wok}, although written in a different form. The $\zeta_2$ terms of tree-level open string amplitudes give the first stringy corrections to gluon scattering, at order ${\cal O}(\alpha^{\prime 2})$, and they also represent the matrix elements of an $N=4$ supersymmetric local operator of the form $\tr(F^4)$.  It has been observed that, up to a constant of proportionality, the $\zeta_2$ terms for MHV amplitudes in string theory coincide with the one-loop all-plus partial amplitudes in gauge theory~\cite{Stieberger:2006bh,Schabinger:2011wh}.  Thus they should obey the same linear relations, and the number of independent quantities should be the same, namely the unsigned Stirling number $c(n-1,3)$.  The number of independent quantities for the $(\zeta_2)^m$ terms in open string amplitudes for $m\geq2$ has been found to be $\sum_{j=0}^m c(n-1,2j+1)$~\cite{Mafra:2021wok}. Interestingly, the ``missing'' $m=1$ member of this sequence is $c(n-1,1)+c(n-1,3) = (n-2)! + c(n-1,3)$, which equals the number of independent one-minus amplitudes from table~\ref{tab:oneminustable}.

Solving the constraints given by eq.~\eqref{dFFrelnsnew} simultaneously, we find that the total number of linearly independent subamplitudes is $c(n-1,3)$ for $4\leq n\leq 8$. Surprisingly, this is exactly the same number found from the single-trace relations \eqref{singletrreln} coming from the $R_0$ matter anomaly cancellation, even though the explicit forms of the relations are distinctly different. Moreover, the number $c(n-1,3)$ is the true number of linearly independent subamplitudes.

From the Casimir color basis, we can actually understand why eq.~\eqref{dFFrelnsnew} gives the unsigned Stirling number of the first kind $c(n-1,3)$ as the number of unconstrained subamplitudes. It turns out that the number of permutations on $n-1$ letters with standard factorization lengths $k$ is exactly equal to $c(n-1,k)$ (see Appendix \ref{sec:factlength}). Of course the total number of permutations on $n-1$ letters is $(n-1)!$. This means that the number of equations appearing in eq.~\eqref{dFFrelnsnew} is
\begin{equation}
  \label{eq:csumformula}
	\sum_{\substack{k=1 \\ k\neq3}}^{n-1}c(n-1,k) = (n-1)! - c(n-1,3).
\end{equation}
Since the $A_n^{dFF}(\sig)$ are linearly independent as formal linear combinations in the subamplitudes $A_n^{[1]}$, a consequence of the fact that the color factors $d_F^{nb_k\dotsc b_1}F^{\sig_kb_k}\cdots F^{\sig_1b_1}$ are also linearly independent for all values of $k$ in $SU(N\geq3)$, and since the $A_n^{[1]}$ appear in $A_n^{dFF}$ with non-zero coefficients, the number of unconstrained subamplitudes remaining after applying \eqn{dFFrelnsnew} is $c(n-1,3)$.

The reason that the single-trace relations \eqref{singletrreln} obtained from the $R_0$ matter cancellation should give the same number of unconstrained subamplitudes as the relations \eqref{dFFrelnsnew} can be understood in the following way. If we take $SU(3)$ as our gauge group, then it turns out that the antisymmetric-tensor representation is isomorphic to the anti-fundamental representation. So, the representation $R_0$ reduces to
\begin{equation}
	\label{R0N=3}
	R_0^{N=3} = 9\,(F\oplus\bar{F}).
\end{equation}
In $SU(3)$, there is no independent $m$-th order Casimir for $m\geq3$. In particular, the quartic Casimir in the fundamental representation is related to the quadratic one as $\tr(X^4) = \frac{1}{2}[\tr(X^2)]^2$, as stated earlier. Therefore cancelling the anomaly by including matter in $R_0^{N=3}$ is equivalent to cancelling the anomaly by including the axion. As a consistency check, the Casimir relation
\begin{equation}
  \lam_G^2\bigl[ \tr(X^2) \bigr]^2 = \tr_G(X^4)
  = \tr_{R_0^{N=3}}(X^4) = 18\,\tr(X^4)
\label{SU3check}
\end{equation}
gives $\lam_G^2=9$, which agrees with eq.~\eqref{lamG}.

\begin{table}[tbp]
	\centering
	\begin{tabular}{|c|c|c|c|c|c|}
		\hline
		& $n=4$ & 5 & 6 & 7 & 8 \\
		\hline 
		$\# S_n/R\mathbb{Z}_n$ & 3 & 12 & 60 & 360 & 2520
		\\
		$c=3$ & 3 & 12 & 59 & 345 & 2344
		\\
		$c=4$ & 3 & 12 & 59 & 345 & 2344
		\\
		$c=5$ & 3 & 12 & 60 & 360 & 2383
		\\
		$A_n^{dFF}(\sigma)$ & 1 & 6 & 35 & 225 & 1624
		\\
		LI & 1 & 6 & 35 & 225 & 1624
		\\
		$c(n-1,3)$ & 1 & 6 & 35 & 225 & 1624
		\\
		\hline
	\end{tabular}
	\caption{The number of unconstrained all-plus YM $n$-point partial amplitudes after applying the constraints \eqref{AFFrelns} and \eqref{dFFrelnsnew}. The variable $c$ refers to the value of $c$ in the double-trace relations~\eqref{AFFrelns}, while $A_n^{dFF}(\sigma)$ refers to imposing eqs.~\eqref{dFFrelnsnew}.}
\label{tab:AFFtable}
\end{table}

Solving the relations \eqref{AFFrelns}, we find that they are more constraining than the double-trace relations \eqref{dbtrreln}, for each value of $c$ except for $c=4$. Table \ref{tab:AFFtable} summarizes the results for $c=3,4,5$ and $4\leq n\leq 8$. The corresponding results for relations \eqref{dbtrreln} were reported in Table \ref{tab:allplustable}. For $c=3$ and $n\geq6$, the relations \eqref{AFFrelns} are non-trivial, and they appear to be equivalent to the $c=4$ case of \eqref{dbtrreln}. Moreover, the cases $c=3,4$ for \eqref{AFFrelns} leave the same number of unconstrained amplitudes as the three-photon-vanishing relations \eqref{dbtrreln} with $c=4$. At $n=8$, the $c=5$ case of \eqref{AFFrelns} is also more constraining than the $c=5$ case of eq.~\eqref{dbtrreln}, but it is not as constraining as the $c=3,4$ cases. However, no choice of $c$ is as constraining as simultaneously solving the relations \eqref{singletrreln} or \eqref{dFFrelnsnew}, which, again, give the true number of linearly independent subamplitudes. 

\section{Conclusions}
\label{sec:conclusions}

In this paper, we explored the consequences of curing anomalies that appear in the twistor uplift of self-dual Yang-Mills theory. In particular, we find that the vanishing of the one-loop all-plus amplitude in these anomaly-free theories leads to various linear relations among the YM all-plus subamplitudes. The one-minus amplitudes do not vanish in these theories, but they are quite simple, and they provide a set of powerful constraints in this case as well.

The inclusion of fermionic matter in the $R_0$ representation \eqref{R0} gives both single-trace relations \eqref{singletrreln} and double-trace relations \eqref{dbtrreln} for the all-plus amplitude. We have shown that the single-trace relations are exactly the same as the ones conjectured to hold in ref.~\cite{Bjerrum-Bohr:2011jrh}. The double-trace relations appear here for the first time, and they include the previously-known three-photon-vanishing relations \cite{Bern:1993qk,Boels:2011tp}. We also argued that the same double-trace relations hold for the one-minus amplitude.

The one-loop amplitude also vanishes in a theory when the axion, described by eq.~\eqref{axionaction}, is included. This requires the gauge group to lack an independent fourth-order Casimir as in eqs.~\eqref{casimiraxion} and \eqref{casimirid}, which is the case for $SU(2)$ and $SU(3)$. In this theory, the all-plus amplitude is cancelled by the tree-level gluon amplitude with a single internal axion exchange. These axion amplitudes naturally have a double-trace structure, or equivalently a double-comb structure. So the Casimir identity \eqref{casimirid} must be used to relate the all-plus amplitude in the DDM basis to the double-comb color structure of the axion amplitude \eqref{fullaxamp}. We utilized the symmetrized-trace (Casimir) decomposition of ref.~\cite{Bandiera:2020aqn} to do this, which converts a trace of Lie algebra generators to a sum on products of symmetrized traces contracted with strings of structure constants. In doing so, we are able to find explicit relations \eqref{dFFrelns} and \eqref{AFFrelns} for the all-plus subamplitudes.

The relations serve to constrain the subamplitudes, giving proposed sizes of the set of linearly independent subamplitudes. We found that the most constraining relations are eqs.~\eqref{singletrreln} and \eqref{dFFrelns}. For $n$-point processes, they both give the unsigned Stirling number $c(n-1,3)$ as the number of unconstrained subamplitudes. This number happens to agree with the true number of linearly independent subamplitudes, as determined from a momentum-twistor parametrization. It is at first surprising that eqs.~\eqref{singletrreln} and \eqref{dFFrelns} agree with each other, but the agreement can be understood from the following two facts about $SU(3)$: (1) the representation $R_0$ reduces to eq.~\eqref{R0N=3}, and (2) the quartic Casimir in the fundamental representation is proportional to the square of the quadratic one. So, curing the anomaly with matter or curing it with the axion is equivalent in $SU(3)$, and, therefore, the relations stemming from these two methods should be equivalent.

It would be interesting to know to what extent the double-trace relations \eqref{dbtrreln} hold for more general helicity configurations. Unlike the all-plus and one-minus one-loop QCD amplitudes, the $\text{N}^k$MHV amplitudes are both infrared (IR) and ultraviolet (UV) divergent, and they generically contain transcendental functions, like polylogarithms. However, they also contain rational terms in their finite remainder, which are the closest analogues to the finite and rational QCD amplitudes. (The divergent terms can be separated from the finite ones by use of the Catani formula \cite{Catani:1998bh}, which predicts the IR divergent terms of UV-renormalized one-loop amplitudes.)  Furthermore, the rational parts of general-helicity $n$-gluon amplitudes are known to obey three-photon vanishing identities~\cite{Boels:2011tp}, which are a subset of the relations obeyed by the all-plus and one-minus amplitudes.  Perhaps a larger subset of these relations also holds.  There may be other relations that include suitable simpler inhomogeneous terms, as in \eqn{AGR} for the one-minus case. We do not expect any such relations among the finite and rational terms to automatically hold for the transcendental terms; e.g.~the one-loop MHV and NMHV 6-photon amplitudes have vanishing rational terms but non-vanishing transcendental terms~\cite{Binoth:2006hk,Binoth:2007ca}.

Another interesting avenue is to understand how relations among one-loop subamplitudes affect the analytic structure at two loops.  For example, the vanishing of the one-loop all-plus amplitude for the $R_0$ theory suggests that the two-loop all-plus amplitude should be finite and rational, when one considers the possible four-dimensional unitarity cuts. Indeed, this is what is found for any number of identical-helicity gluons by using the celestial chiral algebra bootstrap~\cite{Costello:2023vyy}. However, the two-loop all-plus four-point amplitude in the $R_0$ theory is neither IR finite nor rational, when dimensional regularization is used as the IR regulator~\cite{Dixon:2024mzh}.  That is because the four-point case of eq.~\eqref{singletrreln} only holds at $O(\eps^0)$ in dimensional regularization and it fails at higher orders in $\eps$. Consequently, the IR structure of the dimensionally regulated two-loop result, as predicted by the Catani formula \cite{Catani:1998bh}, gives $1/\eps$ poles and transcendental functions. To remedy this, one can compare the rational results of ref.~\cite{Costello:2023vyy} to the IR-subtracted finite remainder, and they do agree for $n=4$~\cite{Dixon:2024mzh}.  Another way to obtain agreement for $n=4$ is to employ a particular mass regulator instead of dimensional regularization~\cite{Dixon:2024mzh}.

Armed with this understanding of IR divergences, it should be possible to utilize relations of the kind discussed in this paper, in order to simplify the computations of new contributions to two-loop all-plus amplitudes for $n>4$.  We look forward to further developments in this direction.

\acknowledgments
We thank Kevin Costello, Song He, Henrik Johansson, Carlos Mafra, and Oliver Schlotterer for helpful discussions. We are particularly grateful to Kevin Costello for hosting us at the Perimeter Institute for Theoretical Physics (PI), where part of this work was completed. This research was supported in part by the US Department of Energy under contract DE--AC02--76SF00515, and in part by PI. Research at PI is supported by the Government of Canada through the Department of Innovation, Science and Economic Development and by the Province of Ontario through the Ministry of Colleges and Universities.

\appendix

\section{Momentum twistor parametrization}
\label{sec:twistorparam}
We briefly summarize momentum-twistor parametrizations of the type given in ref.~\cite{Badger:2013gxa}.

Analytic computations of scattering amplitudes often involve complicated functions of spinor products $\spa{i}.{j}$ and $\spb{i}.{j}$. These quantities are not independent due to non-linear kinematic and algebraic constraints such as momentum conservation and the Schouten identity. Using momentum twistors~\cite{Hodges:2009hk} makes these constraints manifest, allowing for more straightforward simplifications and verifications of identities.

The momentum twistor 
\begin{align}
	Z_i & \equiv 
	\begin{pmatrix}
		\lambda_i \\
		\mu_i
	\end{pmatrix}
\end{align} 
is assigned to particle $i$, where $\lambda_i$ is the usual holomorphic spinor, and the anti-holomorphic spinor $\tilde{\lambda}_i$ is given by
\begin{equation}
	\tilde{\lambda}_i = \frac{\spa{i,}.{i+1}\mu_{i-1} + \spa{i+1,}.{i-1}\mu_i + \spa{i-1,}.{i}\mu_{i+1}}{\spa{i,}.{i+1}\spa{i-1,}.{i}}.
\end{equation}

The momentum twistors transform under the Poincar\'e group, and each particle has a $U(1)$ symmetry: $\lambda_i\to e^{i\theta_i}\lambda_i$ and 
$\mu_i\to e^{i\theta_i}\mu_i$. We can deduce the number of independent kinematic variables from these symmetries. For $n$ particles, there are $4n$ momentum twistor components. The 10 generators of the Poincar\'e group and the $U(1)$ symmetry for each particle reduces the number of independent variables to $4n-10-n=3n-10$. 

At four points, the three kinematic variables are the Mandelstams $s$, $t$, and $u$, which satisfy $s+t+u=0$, giving two independent ones. One parametrization for the momentum twistors is
\begin{align}
	Z = 
	\begin{pmatrix}
		1 & 0 & \frac{1}{s} & \frac{1}{s} + \frac{1}{t} \\
		0 & 1 & 1 & 1 \\
		0 & 0 & 1 & 0 \\
		0 & 0 & 0 & 1
	\end{pmatrix} \,.
\end{align}
For $n\geq5$, a useful parametrization is 
\begin{align}
	\label{twistparam}
	Z = 
	\begin{pmatrix}
		1 & 0 & y_1 & y_2 & y_3 & y_4 & \dots & y_{n-3} & y_{n-2} \\
		0 & 1 & 1 & 1 & 1 & 1 & \dots & 1 & 1\\
		0 & 0 & 0 & \frac{x_{n-1}}{x_2} & x_n & x_{n+2} & \dots & x_{3n-12} & 1 \\
		0 & 0 & 1 & 1 & x_{n+1} & x_{n+3} & \dots & x_{3n-11} & 1 - \frac{x_{3n-10}}{x_{n-1}}
	\end{pmatrix} \,,
\end{align}
where the $x_i$ are the $3n-10$ independent kinematic variables and $y_i \equiv \sum_{j=1}^i\prod_{k=1}^{j}x_k^{-1}$.

\section{Standard factorization lengths and unsigned Stirling numbers}
\label{sec:factlength}

Let $c(n,k)$ be an unsigned Stirling number of the first kind, and let $\ell(n,k)$ denote the number of permutations with a standard-factorization length of $k$, i.e.
\begin{equation}
	\ell(n,k) \equiv \#\{\sig\in S_n|\sig=\sig_k\cdots\sig_1\}.
\end{equation}
We will show that $c(n,k)=\ell(n,k)$, by showing that they satisfy the same recurrence relations
\begin{equation}
	\label{lenrecurreln}
	\ell(n+1,k+1) = n\cdot\ell(n,k+1) + \ell(n,k)
\end{equation}
with $\ell(0,0) = 1$ and $\ell(n,0)=\ell(0,n)=0$ for $n>0$.

Recall that $\sig=\sig_k\sig_{k-1}\cdots\sig_1$ denotes the standard factorization of a permutation (or word) $\sig\in S_n$ into subwords $\sig_1,\dotsc,\sigma_k$ with the properties:
\begin{enumerate}[(1)]
	\item The first letter of $\sig_i$ is greater than the first letter of $\sig_j$ whenever $i>j$.
	\item The first letter of $\sig_i$ is the minimum letter appearing in $\sig_i$.
\end{enumerate}
This factorization is unique, and $\sig$ is viewed as the concatenation of these subwords.

Consider $\ell(n+1,k+1)$. Every permutation of $S_{n+1}$ can be built from a permutation $\sig\in S_n$ by placing $n+1$ either before or after an element in $\sig$. Consider placing $n+1$ after $\sig(i)$ for $1\leq i\leq n$ like so
\begin{equation}
	(\sig(1),\dotsc,\sig(i),n+1,\sig(i+1),\dotsc,\sig(n))\in S_{n+1},
\end{equation}
where it is understood that nothing appears after $n+1$ if $i=n$. Since $1\leq\sig(i)\leq n<n+1$ for all $1\leq i\leq n$, the placement of $n+1$ does not affect the factorization length. The number of such permutations with length $k+1$ is then $n\cdot\ell(n,k+1)$. Placing $n+1$ at the front of $\sig$, i.e. before $\sig(1)$, will always increase the factorization length, since $n+1>\sig(1)$. So, the number of these types of permutations in $S_{n+1}$ with length $k+1$ is $\ell(n,k)$. The initial conditions are trivially satisfied.

Next, we show that the unsigned Stirling numbers satisfy the same relation. Recall that $c(n,k)$ is the number of ways to partition a set of $n$ elements into $k$ distinct, cyclicly ordered sublists, or, in terms of cycle notation, the number of permutations in $S_n$ that can be written as the product of $k$ disjoint cycles. From these definitions, the initial conditions $c(0,0)=1$ and $c(n,0)=c(0,n)=0$ are trivially satisfied.

Consider $c(n+1,k+1)$, and distinguish $n+1$ as before. Starting with $k+1$ cycles built from the set $\{1,2,\dotsc,n\}$, we can place $n+1$ after the element $i$ within the same cycle as $i$. There are $n$ ways to do this. So, the number of partitions of $\{1,\dotsc,n,n+1\}$ with $n+1$ sharing a cycle with another element is $n\cdot c(n,k+1)$. The only other option for placing $n+1$ is to place it into a cycle by itself, increasing the number of cycles by one. There are $c(n,k)$ such partitions. Thus,
\begin{equation}
	c(n+1,k+1) = n\cdot c(n,k+1) + c(n,k).
\label{eq:crecurse}
\end{equation}

A generating function for the unsigned Stirling numbers is:
\begin{equation}
  x (x+1) \cdots (x+n-1) = \sum_{k=0}^n c(n,k) \, x^k \,,
  \label{Stirlinggenfn}
\end{equation}
which leads to the recursion relation~\eqref{eq:crecurse}, as well as
to the correct boundary conditions.




\bibliographystyle{JHEP}
\bibliography{relations.bib}

\end{document}